\documentclass[twocolumn]{aastex631}

\usepackage[T1]{fontenc}
\accepted{to The Astrophysical Journal}

\DeclareRobustCommand{\VAN}[3]{#2}
\let\VANthebibliography\thebibliography
\def\thebibliography{\DeclareRobustCommand{\VAN}[3]{##3}\VANthebibliography}

\usepackage{graphicx}	
\usepackage{amsmath}	

\usepackage{dirtytalk}

\graphicspath{{./}{figures/}}

\begin{document}

\title{Co-Evolution of Stars and Gas: Using Analysis of Synthetic Observations to Investigate the Star-Gas Correlation in STARFORGE}

\author{Samuel Millstone}
\affiliation{Department of Astronomy, University of Massachusetts Amherst, Amherst, MA 01003, USA}

\author{Robert Gutermuth}
\affiliation{Department of Astronomy, University of Massachusetts Amherst, Amherst, MA 01003, USA}

\author{Stella S. R. Offner}
\affiliation{Astronomy Department, The University of Texas at Austin, Austin, TX 78712, USA}

\author{Riwaj Pokhrel}
\affiliation{Department of Physics and Astronomy, The University of Toledo, Toledo, OH 43606, USA}

\author{Michael Y. Grudi\'{c}}
\affiliation{Carnegie Observatories, 813 Santa Barbara St, Pasadena, CA 91101, USA}



\begin{abstract}

We explore the relationship between stellar surface density and gas surface density (the star-gas or S-G correlation) in a 20,000 M$_{\odot}$ simulation from the STAR FORmation in Gaseous Environments ({\sc starforge}) Project. We create synthetic observations based on the \textit{Spitzer} and \textit{Herschel} telescopes by modeling active galactic nuclei contamination, smoothing based on angular resolution, cropping the field-of-view, and removing close neighbors and low-mass sources. We extract star-gas properties such as the dense gas mass fraction, the Class II:I ratio, and the S-G correlation ($\Sigma_{\rm YSO}/\Sigma_{\rm gas}$) from the simulation and compare them to observations of giant molecular clouds, young clusters, and star-forming regions, as well as to analytical models. We find that the simulation reproduces trends in the counts of young stellar objects and the median slope of the S-G correlation. This implies that the S-G correlation is not simply the result of observational biases but is in fact a real effect. However, other statistics, such as the Class II:I ratio and dense gas mass fraction, do not always match observed equivalents in nearby clouds. This motivates further observations covering the full simulation age range and more realistic modeling of cloud formation.

\end{abstract}





\section{Introduction}
\label{section:Section 1}

The majority of stars form in associations or groups within giant molecular clouds \citep[GMCs,][]{Ladea91, Kruea19, Cheea22}, which can vary greatly in size, from $\sim$10 to thousands of stars \citep{Porea04}. Feedback from embedded clusters often quickly disperses the natal clump or even the entire GMC \citep{Lad05, Kraea20}. Therefore, the relationship between gas and young stellar object (YSO) density provides important clues about the star formation process and cloud evolution. \cite{Sch59} was one of the first to present an analytical model of the relationship between star formation rate (SFR), and thus stellar mass, and gas density. That work suggested that SFR and gas density follow a power law relationship.

This correlation was examined over the next several decades by a number of authors \citep[e.g.][]{San69, Har71}. However, it was not until improved observational capabilities and analysis techniques in the 1980s and 1990s \citep[e.g.][]{Ken89, Ken98} that strong evidence was found for its viability. This work motivated an analogous relation known as the Kennicutt-Schmidt (KS) law that applies to line-of-sight surface densities of gas and the star formation rate per area:

\begin{equation}
     \Sigma_{\text{SFR}} \propto \Sigma_{\text{gas}}^N.
\end{equation}

Henceforth, we refer to this relation as the Star-Gas or S-G correlation. This relationship has since been well-characterized as a power-law with an index of N$\sim$1.4 as applied to galaxy-scale star formation (see \citet{KE12} for a detailed review). 

At smaller scales within individual galaxies, there is also evidence for the presence of an S-G correlation. For example, \cite{Bigea08} used HI, CO, 24 $\mu$m, and UV data to examine the S-G correlation at 750~pc resolution in 18 nearby spiral and dwarf galaxies. Many regions showed a strong power-law relation, although the power-law index varied from 1.1 to 2.7 based on position. They also observed that the star formation efficiency (SFE) decreased with galactic radius, which they argued implies a connection between environment and the S-G correlation.

However, the methods used to measure the SFR on $\gtrsim$ kpc scales, such as H$\alpha$, far-UV, and 24 $\mu$m emission, become less effective at smaller spacial scales. The results of \cite{Liuea11}, as well as modeling by \cite{CLK12} show that this kind of analysis breaks down with shrinking sample area because star formation is not well-sampled statistically. \cite{Gutea11} (G11 hereafter) demonstrated that the SFR calculated from far-IR luminosity \citep[$L$\textsubscript{FIR}, e.g.,][]{Heiea10} underestimates the SFR calculated from counts of YSOs in nearby young clusters by up to an order of magnitude. This is because measurements based on far-IR luminosity assume a well-sampled stellar initial mass function (IMF) and reliable sampling of the GMC mass function to fully sample the lifetime of high-mass stars. However, in order to satisfy these assumptions, measurements must be integrated over physical scales $\gtrsim 1$ kpc \citep{CLK12}.

To avoid the smoothing inherent to measurements of star formation relations in other galaxies, some recent studies instead focus on individual star-forming regions in the local Milky Way, where it is possible to identify and count individual forming stars with high completeness. Since YSOs provide a direct measurement of the SFR, a simple estimate of the total mass converted to stars per time is given by

\begin{equation} 
\dot{M}=\frac{m_{\rm YSO}{n_{\rm YSO}}}{t_{\rm avg}},
\end{equation}

where $m_{\rm YSO}$ is the average mass of a YSO, $n_{\rm YSO}$ is the number of YSOs, and $t_{\rm avg}$ is the characteristic timescale for the YSO evolutionary stage or stages considered. 

By utilizing YSO censuses from \textit{Spitzer}, G11 and \cite{Pokea20} (P20 hereafter) found and measured an intracloud S-G correlation with an index of N\,$\approx$\,2 in several nearby GMCs. While initial measurements varied widely (N\,=\,1.5~-~4) (G11), P20 reduced intrinsic scatter in the measurements by adopting a uniform YSO extraction from the {\it Spitzer} Extended Solar Neighbor Archive (SESNA), utilizing more robust \textit{Herschel}-based GMC gas column density maps, and by specifically using YSOs in the early stages of star formation. This led to N\,=\,1.8~-~2.3 in 12 nearby clouds with gas masses varying over three orders of magnitude. Also, the scaling factor in the S-G correlation varies between clouds \citep[G11, P20]{Ladea13}, but the scatter in the scaling factor is reduced significantly when it is normalized by the gas freefall time \citep{Pokea21}. This implies that the SFE per freefall time has limited variation, which may indicate that local processes (e.g., protostellar outflows and stellar winds) govern and regulate star formation \citep{Guszejnov2021IMF, Pokea21, Huea22}. 

In order to gain a better understanding of how local processes impact star formation, it is useful to turn to theoretical models and numerical simulations. However, observed S-G correlations have only recently started to become incorporated as constraints for models of star-forming molecular gas. P20 used simulations by \cite{Qiaea15} that used the {\sc ORION} adaptive mesh refinement code \citep{Truea98, Klein99} to create synthetic observations similar to observations taken by {\it Herschel}. That work reproduced similar S-G correlations for 12 nearby GMCs using hydrodynamic turbulent simulations and an analytical model of thermal fragmentation. While the simulation produced an S-G correlation that is very similar to observations, it did not include magnetic fields or kinematic feedback. In this work, we analyze a $20,000$ M\textsubscript{$\odot$} run of the STAR FORmation in Gaseous Environments ({\sc starforge}) project, the first massive GMC magnetohydrodynamics simulation to resolve individual stars while including multi-band radiation, stellar winds, protostellar outflows, and supernovae \citep[etc.]{Gruea21, Gruea22}.

In order to most effectively compare the {\sc starforge} simulation to observations, we construct synthetic observations according to the data used in P20, taking into account the known specifications and limitations of \textit{Spitzer} and \textit{Herschel} data. In Section \ref{section:Section 2}, we describe the specifics of the simulation snapshots and our methods for creating synthetic observations. In Section \ref{section:Section 3}, we present results from our investigation into various star-gas properties in the simulation and compare to observations. Discussion is provided in Section \ref{section:Section 4}, and a summary and conclusions are given in Section \ref{section:Section 5}.

\section{METHODS}
\label{section:Section 2}

\subsection{{\sc starforge} Simulations}

The {\sc starforge} framework is built on the {\sc gizmo} meshless finite mass magnetohydrodynamics code \citep{Hopkins2015}. The framework includes a variety of modifications that enable the modeling of individual forming stars and the interactions that occur with their cloud environment. In this work we analyze the {\sc starforge} simulation presented in \citet{Gruea22}. We briefly summarize the simulation properties here and refer the reader to \citet{Gruea21} for a detailed description of the {\sc starforge} numerical methods. 

The simulation follows the evolution of a 20,000 M$_{\odot}$ cloud with initial radius of 10~pc. The cloud turbulence is initialized so that the cloud is virialized with $\alpha \equiv 5 \sigma_{\rm 3D }^2 R_{\rm cloud} / (3GM_{\rm cloud}) = 2$, where $\sigma_{\rm 3D}$ is the gas velocity dispersion. The initial magnetic field is uniform in the $\hat z$ direction and corresponds to a mass-to-flux ratio relative to the critical value for stability $\mu \equiv 0.4 \sqrt{E_{\rm grav}/E_{\rm mag}} = 4.2$, where $E_{\rm grav}$ and $E_{\rm mag}$ are the total gravitational and magnetic energies, respectively. 

The calculation follows the gas thermodynamics self-consistently, including treatment of line cooling, cosmic-ray heating, dust cooling and heating, photoelectric heating, hydrogen photoionization, and collisional excitation of both hydrogen and helium. The evolution of the dust temperature is coupled to the radiative transfer step. {\sc gizmo}'s radiation transfer module follows five bands, which cover the frequencies corresponding to ionizating radiation, FUV, NUV, optical-NIR, and FIR \citep{HopkinsGrudic2019,Hopkins2020}.

Once gas satisfies multiple criteria intended to identify centers of unstable collapse, Lagrangian sink particles are inserted, which occurs at densities of $\rho_{\rm max}\sim 10^{-14}$\,g\,cm$^{-3}$. The cell mass resolution is $dm=10^{-3}\,$M$_\odot$, which allows the calculation to resolve the stellar mass spectrum down to $\sim 0.1\,$M$_\odot$. The sink particles, henceforth referred to as stars, follow a sub-grid model for protostellar evolution and radiative feedback as described in \citet{Offner2009}. The particles are also coupled to models describing protostellar outflow launching, stellar winds, and supernovae \citep{Cunningham2011,Guszejnov2021IMF,Gruea21}. The calculation continues until stellar feedback disperses the natal cloud and star formation concludes, which happens at $\sim 9$\,Myr.

The simulation has a final SFE of $8$\% that agrees with statistical models of nearby galaxies. Protostellar jets dominate feedback for most of the simulation and are important for regulating the IMF, but they cannot wholly disrupt the cloud. Eventually, radiation and winds from massive stars create bubbles that expand and disrupt the cloud, drastically reducing SF. By following GMC evolution, \cite{Gruea22} measure a relatively unambiguous IMF. It resembles the Chabrier IMF with a high-mass slope of $\alpha=-2\pm0.1$. The IMF is much more realistic than previous simulations without full feedback. Feedback from radiation/winds of massive stars limits the maximum observed mass to 55 solar masses, moderating the high-mass tail of the IMF. The integrated luminosity and ionizing photon rate are also very close to an equal-mass cluster with a canonical IMF. A more detailed study of the impact of various feedback processes and cloud initial conditions on the IMF is presented in \cite{Gusea22}. \cite{Gruea22} also note the importance of directly comparing observations and simulations via synthetic observations, as we aim to do in this work.

To construct the stellar surface density, we require a minimum of 11 YSOs. The first snapshot with at least this number of sources is at 1.47 Myr. Altogether our analysis uses 16 snapshots, spaced 0.49 Myr apart, which span 1.47 to 8.80 Myr.

\subsection{Constructing Synthetic Observations}
\label{section:Constructing Synthetic Observations}

For our analysis to better mirror that of P20, we create synthetic observations by including various considerations to bring our data closer to that which might have been observed by \textit{Spitzer} and \textit{Herschel}. We refer to analysis done with minimal adjustments, i.e., only 2D projection, age-to-class conversion, and 0.01 M$_{\odot}$ mass cutoff (see below) as the ``fiducial analysis'', while analysis with further considerations are collectively referred to as ``synthetic observations.'' The fiducial (minimally-adjusted) case allows us to examine how well the simulation can reproduce various statistics and identify where observational biases may affect the agreement. In order to create these synthetic observations, we extract or compute the (line-of-sight-projected when applicable) molecular number density of H\textsubscript{2} and the masses, coordinates, ages, and particle indexes of the sink particles, which represent YSOs.

\subsubsection{YSOs}

YSOs fall into distinct groups based on their observed properties. Historically, these have been binned into representative classes \citep{Lad87, SAL87, Greea94, Robea07, Dunea15}, e.g., Class I, Class II, and Class III.\footnote{There are other YSO classes, such as Class 0 and Flat spectrum, but these protostellar sub-types are not differentiated in our adopted data, and are thus not considered in our analysis.}. Note that class does not have a direct mapping to source age, but it is often used as a proxy for evolutionary stage. YSOs in each class differ in the shape of their spectral energy distribution (SED), which depends on the characteristics of the circumstellar material around the YSO. Class Is are usually deeply embedded in cold, dense, and dusty gaseous envelopes, Class IIs have classical protoplanetary disks, and Class IIIs have mostly lost their disks (or the visible disk material has substantially coalesced into larger planetesimals that are generally invisible in the infrared).

For the first step of our analysis, we map each of the {\sc starforge} stars to an observational class. Ideally, the stellar age would be employed to directly map each source to the appropriate spectral class. However, the average age and lifetime of each class is uncertain, since the individual classes are not completely distinct and the boundaries between them are somewhat arbitrary. Class lifetimes are inferred observationally using the relative number of sources in each class and by assuming a typical disk lifetime \citep[e.g.][]{Dunea14}. Consequently, a full self-consistent class assignment requires constructing synthetic observations using radiative transfer to model the SEDs. Instead, we assign each star to a class based on its age (the time elapsed since the sink particle forms in the simulation) and adopt a statistical approach rather than an exact mapping.

We model the transitions from Class I $\rightarrow$ Class II and Class II $\rightarrow$ Class III as exponential decays, adapting the models and half-lives of the transitions from \cite{KD18} and \cite{Mam09} to represent the age-to-class conversion. Using these half-lives, we calculate two numbers corresponding to each source, $f_{a}$ and $f_{b}$, which corresponds to the statistical weighting given to each source for transitions (a) (Class I $\rightarrow$ II) and (b) (Class II $\rightarrow$ III):

\begin{equation}
f_{a,b}=\textrm{e}^{-t_{\textrm{age}}\frac{{\rm ln}(2)}{t_{1/2;a,b}}},
\end{equation}

where $t_{\textrm{age}}$ is the age of the YSO and $t_{1/2;a}$ and $t_{1/2;b}$ are the half-lives of the Class I $\rightarrow$ Class II transition (0.22 Myr) and the Class II $\rightarrow$ Class III transition (1.7 Myr), respectively. Then, we generate two random numbers ($r_{a}$ and $r_{b}$) for each source using consistent seeds and the persistent source index from {\sc starforge}, so that each YSO has the same $r_{a}$ and $r_{b}$ for the entire run. If $r_{a} < f_{a}$, then the YSO is assigned to Class I. If not, we check whether $r_{b} < f_{b}$. If so, the YSO is assigned to Class II. If not, it is assigned to Class III.

By fixing $r_{a}$ and $r_{b}$ for each source, we ensure that the sources progress forward through the classes (I to II to III) as they age in the simulation. However, in actual observations, a YSO's trajectory may not be so linear. For example, \cite{Dunea10} used models of accreting sources to show that YSOs undergoing episodic mass accretion may transition to an earlier Class. The notion that older sources can populate the earlier classes is also supported by the work of \cite{Herea07}, who observed what appear to be older, ``evolved'' disks. Another problematic assumption is that the Class lifetimes are the same for every environment, which is unlikely since protostars in areas of high YSO density tend to have greater luminosity \citep{Kry14, Cheea22}. Despite the approximate nature of our model for Class assignment, we find that it reproduces the expected YSO distributions well. Whereas, assuming an exact one-to-one mapping between age and Class leads to sharp transitions that do not match observations as closely.

\begin{figure*}%
    \centering
    \includegraphics[width=\textwidth]{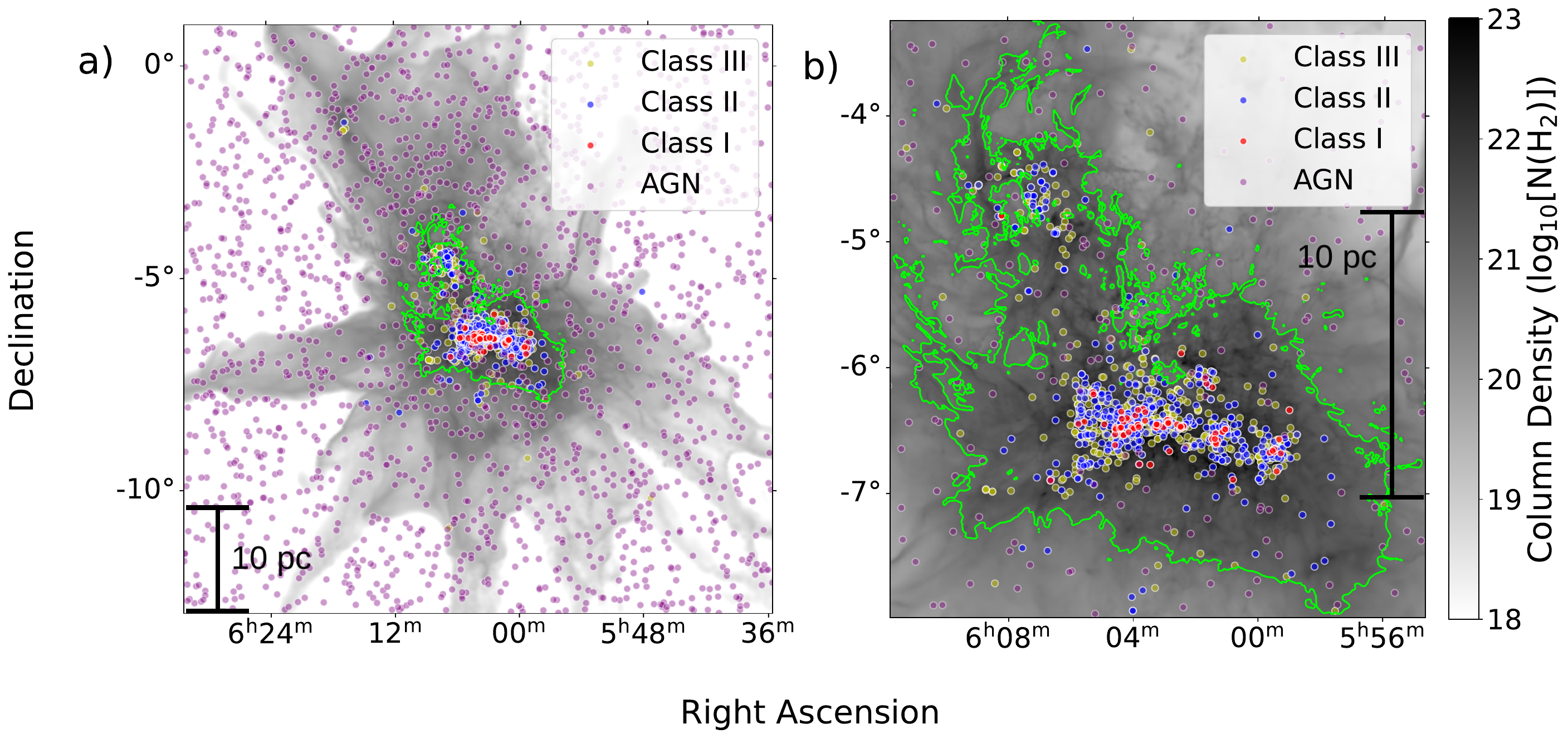}%
    \caption{Projected N(H\textsubscript{2}) column density map of a 200 pc-distance cloud with N(H$_{2}) = 10^{21}$~cm$^{-2}$ contour over-plotted in green. Colored circles indicate the locations of YSOs and AGN. a) Full field of view of the simulation at $\sim$5.4 Myr. b) Zoomed ($\sim$20-pc) field of view cropped to the furthest extent of the green contour at $\sim$5.4 Myr. AGN contaminants dominate the source counts in the low-column density regions.}
    \label{fig:AGN Example}%
\end{figure*}

Next, in order to model source confusion present in \textit{Spitzer} observations, we inject Active Galactic Nuclei (AGN) contaminants. In \textit{Spitzer} observations, background AGN can appear as YSOs of Class I and II with roughly equal probability \citep{Gutea08,Gutea09}. To simulate this effect, we randomly place $N$ Class Is and IIs within the dataset, where $N$ was determined to be $\sim$9 per square degree (P20). This has the immediate effect of introducing many sources with low spatial density. This is especially significant for the synthetic clouds at closer distances due to the commensurately larger angular size of the cloud (see Figure \ref{fig:AGN Example}, where it is clear that AGN dominate over YSOs in low gas density regions). We then correct for these contaminants following the method used by G11: we adopt a threshold of log $\Sigma_{\rm gas}$ $>$ 1.3 M$_{\odot}$ pc$^{-2}$ for points on the S-G plot (see Section \ref{section:Demonstration of Effects}). We adopt the same distribution of AGN contaminants for all snapshots to ensure that the AGN stay the same (i.e. same position and Class).

We also model instrumental detection limits to account for undetectable low-luminosity sources. To replicate this in the synthetic observations, we implement a simple mass cutoff, where we remove sources below 0.08 M$_{\odot}$ (200 and 400 pc distance) or 0.2 M$_{\odot}$ (800 pc).\footnote{Note that we implement a \textit{global} 0.01 M$_{\odot}$ mass cutoff for all of our analysis, including the fiducial case, in order to avoid spurious sources with extremely low masses that were the result of a known bug in that simulation that has since been fixed, eliminating the erroneous sources.}

Last, we model \textit{Spitzer}’s limited angular resolution by removing stars in close proximity. When a source and its nearest neighbor (YSO or AGN) are within the adopted beam size threshold of 5\arcsec, we remove the lower-mass source. We assign AGN a mass of 1.1 M$_{\odot}$ to avoid losing them to the mass cutoff. We do only one pass to remove sources, but this is sufficient to remove the vast majority of close neighbors.

\subsubsection{Gas}

We construct 2D projected column density maps with cloud distances of 200, 400, and 800 pc, which are chosen to model the majority of the clouds in the P20 sample. Figure \ref{fig:AGN Example} shows one of these maps with a spatial distribution plot of YSOs and AGN contaminants.

\begin{figure*}%
    \centering
    \includegraphics[width=\textwidth]{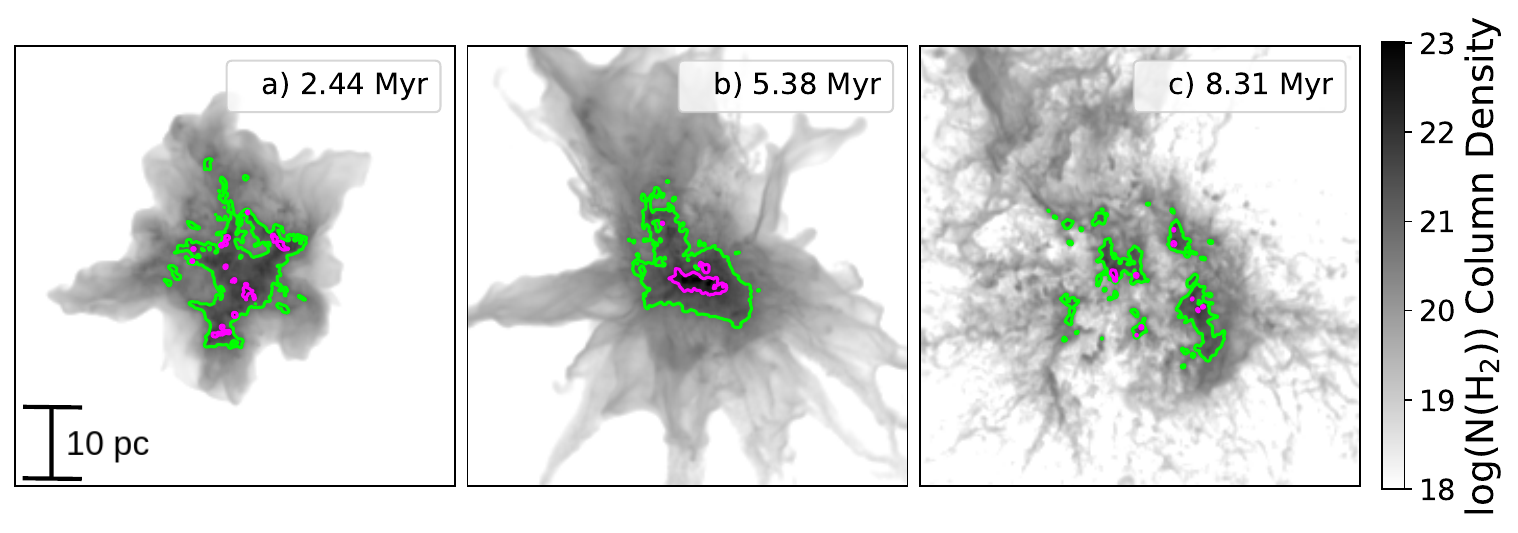}%
    \caption{Projected N(H\textsubscript{2}) column density map with N(H$_{2}) = 10^{21}$~cm$^{-2}$ contour over-plotted in green and N(H$_{2}) = 10^{22}$~cm$^{-2}$ contour in magenta at (a) $\sim$2.4 Myr. (b) $\sim$5.4 Myr. (c) $\sim$8.3 Myr. The green contour outlines the likely \textit{Spitzer} field of view for an equivalent cloud. Note that the high-density (magenta) region coalesces as star formation increases and eventually breaks apart due to stellar feedback, which is in the process of dispersing the cloud in (c).}%
    \label{fig:Spatial Distribution}%
\end{figure*}

The \textit{Spitzer} and \textit{Herschel} fields of view focus on regions of high column density (clumps) within the clouds. To simulate this, we crop the gas maps to the bounds set by a 10\textsuperscript{21} cm$^{-2}$ column density contour on a 120\arcsec-smoothed gas map constructed specifically for this purpose. This map is not used again in the further analysis. We smooth to keep small overdensities from artificially enlarging the cropping area.\footnote{Such overdensities did occasionally occur, see discussion in Section \ref{section:Section 4}. However, the low-density cutoff described in section \ref{subsection:S-G Correlation} reduces the impact of the increased area.} This greatly reduces the field of view compared to the full view, as shown in Figure \ref{fig:Spatial Distribution}, and makes our maps more similar to the \textit{Spitzer} and \textit{Herschel} data we compare with. Additionally, this significantly reduces the amount of low-density AGN contamination (see Figure \ref{fig:AGN Example} and Section \ref{section:Section 4} for more details). In order to simulate the angular resolution of \textit{Herschel}, the gas maps are smoothed with a 36\arcsec \,Gaussian kernel.

\section{ANALYSIS}
\label{section:Section 3}
\subsection{Overview Statistics}

To better compare with observations, We first define a few bulk cloud properties. We define the total cloud gas mass, M$_{\rm gas}$, as the combined mass of gas at column densities $10^{21}$~cm$^{-2}$ and above. Similarly, the dense gas mass, M$_{\rm dense}$, is the total gas at column densities $10^{22}$~cm$^{-2}$ and above. The dense gas mass fraction is then the ratio of dense to cloud gas mass. This metric gives an indication of the fraction of the cloud that is most likely to form clusters \citep{Batea14,Heyea16}. We define the disk fraction as the ratio of the number of Class I and II YSOs to the total number of YSOs regardless of circumstellar material. Disk fraction can be used as a proxy for the population age \citep{Haiea01, Herea07b}. A similar statistic, the Class II to Class I ratio, is generally believed to be a good relative evolution indicator for YSOs, especially for earlier evolution (G11, P20).

\begin{figure}
    \includegraphics[width=0.47\textwidth]{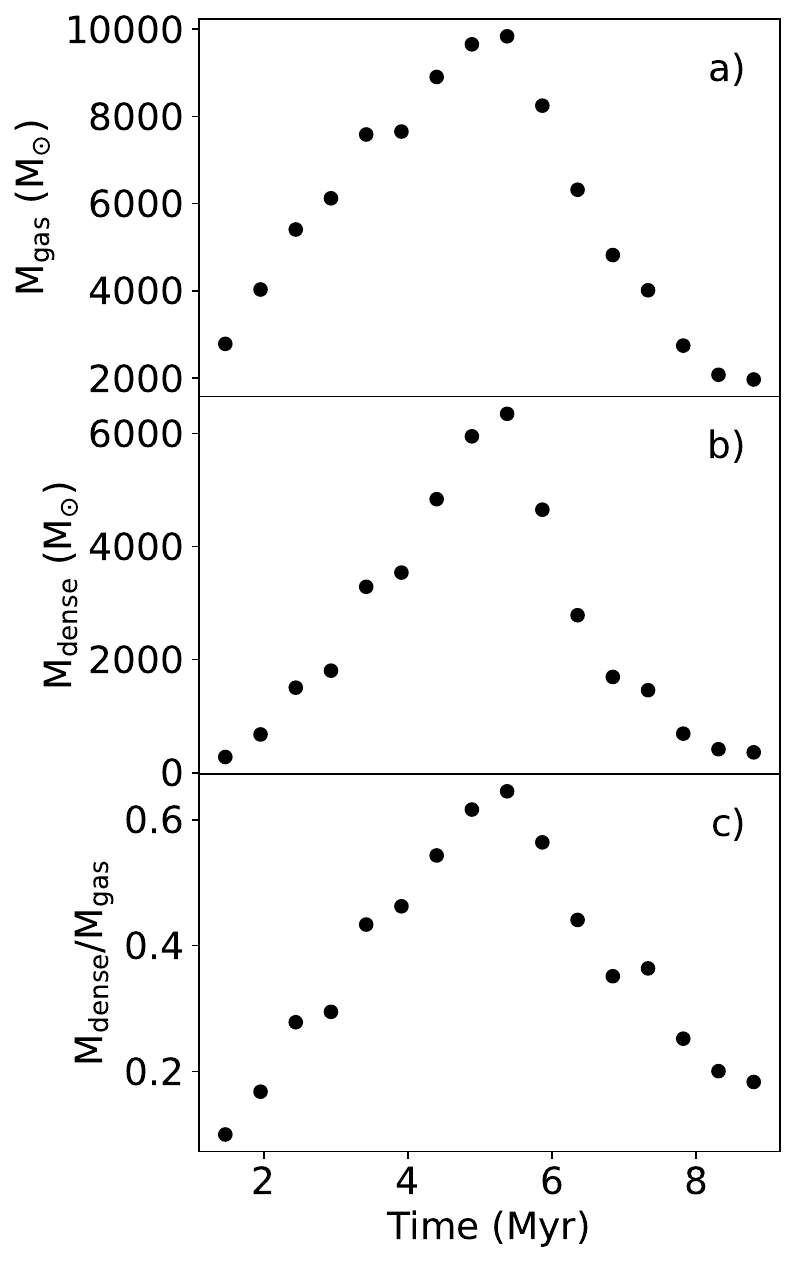}
    \centering
    \caption{Evolution of gas in the {\sc starforge} simulation versus time: (a) Cloud mass at column densities N(H$_{2})>10^{21}$~cm$^{-2}$, (b) Mass of the dense gas at column densities N(H$_{2})>10^{22}$~cm$^{-2}$, (c) Dense gas mass fraction.}
    \centering
    \label{fig:Dense Gas Fraction}
\end{figure}

Figure \ref{fig:Dense Gas Fraction} shows the evolution of the gas properties with time for the fiducial case. The cloud mass and dense gas mass increase steadily over time and peak at $\sim$ 5.4 Myr. The maximum mass reaches about half of the 20,000 M\textsubscript{$\odot$} of gas that makes up the entire simulated GMC. After this point, the cloud mass decreases rapidly to less than 1/10 of the initial GMC mass. The dense gas mass fraction exhibits a similar trend, peaking at $M_{\rm dense}/M_{\rm cloud} \sim 0.6$ at the same time.

\begin{figure}%
    \centering
    \includegraphics[width=0.47\textwidth]{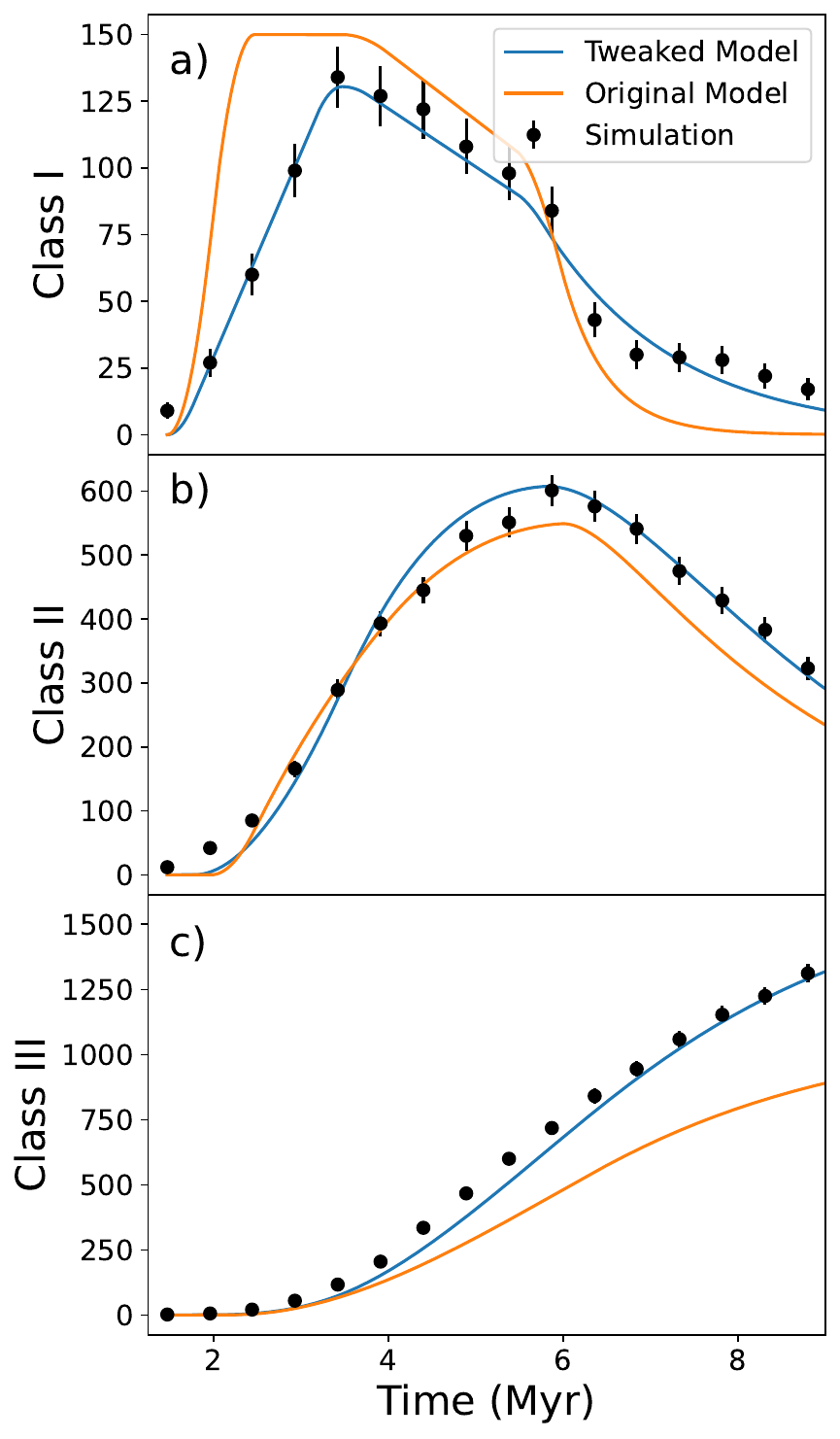}%
    \caption{Evolution of each Class of {\sc starforge}-derived YSO counts in this work (black points) overlaid with analytical models adapted from \citet{MGK22}: (a) Number of Class I sources versus time, (b) Number of Class II sources versus time, (c) Number of Class III sources versus time. The orange lines are shifted and rescaled versions of the \citet{MGK22} models using their parameter selections, while the blue lines adopt parameter value adjustments to achieve strong agreement with the STARFORGE data.}%
    \label{fig:Model Comparison}%
\end{figure}

Figure \ref{fig:Model Comparison} shows that the number of Class I and II sources evolve in a similar way to the gas. Star formation increases steadily for the first 3.43~Myr as indicated by the rising number of Class Is. After 3.43~Myr, star formation declines to 63\% of the maximum in 2.45~Myr and then drops to only half this value in the next snapshot. The number of Class IIs evolves more gradually, peaking at 5.88~Myr, after which point it steadily decreases to about half its maximum value by the end of the simulation.

Figure \ref{fig:Model Comparison} also shows an analytical model adapted from \cite{MGK22} created to predict the populations of Class I, II, and III objects. The original model is semi-empirically developed to generate the ensemble of Class II:I ratios observed in \cite{Gutea09}, and thus we expect it to describe the progression of star formation in nearby clustered regions reasonably well. The idea of a strong central peak in SFR is characteristic of a cluster formation event, which supports the use of a cluster-derived model. It is shown with a vertical stretch and time axis shift to make the model more visible without adjusting the main model parameters.

We also plot a tweaked version of the model with minor adjustments to better fit our assumptions and outputs. Namely, we shift the time axis by 1.47 Myr to match our snapshots, increase the SFR from 100 to 435 (unitless metric which changes the vertical scale of the model), lengthen the rise and decay times for the Class Is from 0.5 to 1.7 Myr and 0.5 to 1.5 Myr, respectively, and shorten the lifetimes of Class Is and IIs to be closer to (but not exactly the same as) the half-lives for our adopted Class transitions (0.5 to 0.3 Myr and 2.0 to 1.5 Myr, respectively). With these parameters, the model reproduces the fiducial {\sc starforge} data remarkably well. This suggests the {\sc starforge} simulation provides a good representation of cluster formation. As we shall see below, the simulation appears to agree less well with star formation observed in full GMCs, which generally contain multiple distinct star-forming regions and have longer and more complex star formation histories.

\begin{figure}%
    \centering
    \includegraphics[width=.47\textwidth]{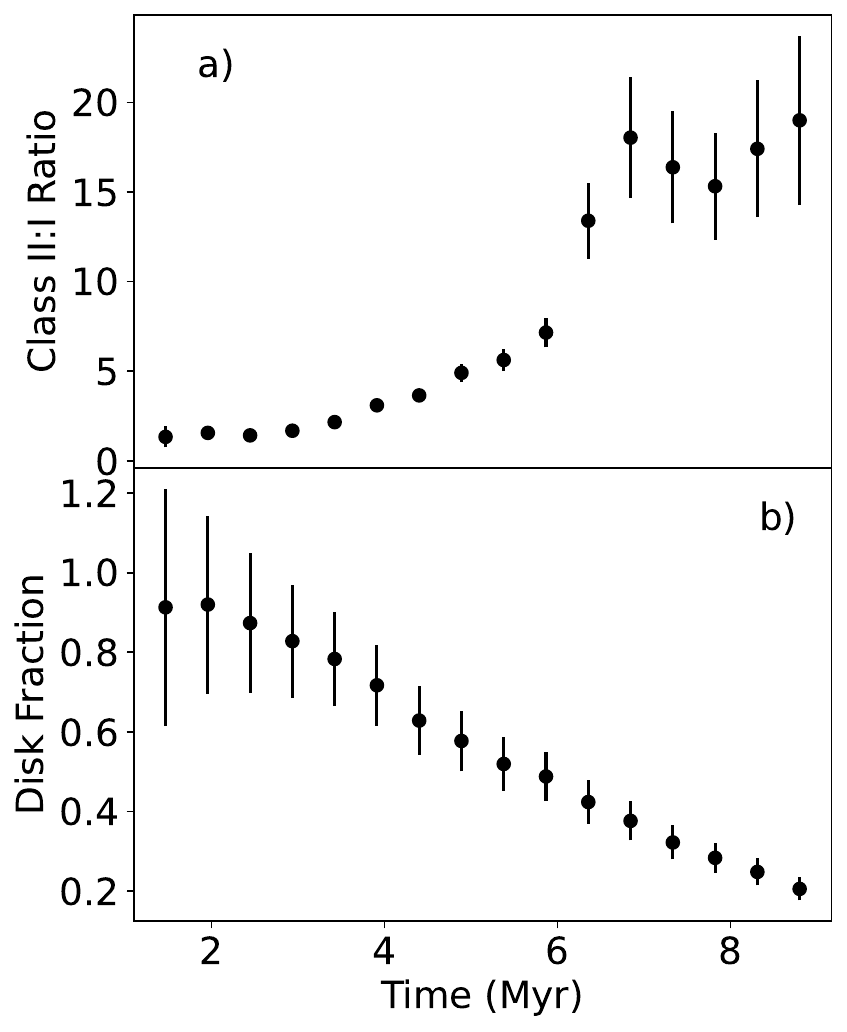}%
    \caption{(a) Class II:I ratio versus time increases steadily until about ~6 Myr, at which point it jumps up and does not follow a consistent trend. (d) Disk fraction, i.e., ratio of Class I and Class II sources to total number of sources, decreases steadily, but slower than comparable observations based on mean stellar ages in real clouds. Error bars are calculated through standard error propagation.}%
    \label{fig:Fig II:I and Disk Fraction}%
\end{figure}

Figure \ref{fig:Fig II:I and Disk Fraction} shows the evolution of the Class II:I ratio and the disk fraction in this simulation. The disk fraction starts near 1.0 and then decreases nearly linearly to 0.21 in the final snapshot. This is more drawn-out than the traditional disk fraction versus stellar age plot \citep[e.g.][]{Mam09}. The {\sc starforge} calculation exhibits a broad range of Class II:I ratios, which span 1.3-19.0. For comparison, P20 recorded the Class II:I ratio and the cloud mass for 12 clouds at distances between 140 and 1400~pc. They found that the Class II:I ratio remained between $\sim$3.5-9.7 for each cloud observed, which is a much narrower range than we find in the {\sc starforge} snapshots. However, the P20 values are uncorrected for AGN and edge-on disk contamination, which would likely change the Class II:I ratios, as will be seen below.

Using publicly available \textit{Herschel} data \citep[][P20]{Andea10}, we calculate the dense gas mass fraction of the clouds and clusters P20 and \citet{Gutea09} observed. We adopt the publicly available YSO lists from SESNA, correct for AGN and edge-on disk contamination, and crop for coverage consistency and to the N(H$_{2}) = 10^{21}$~cm$^{-2}$ limit. In the case of the \citet{Gutea09} data shown, we adopt all ``cluster cores'' that overlap with clouds from the P20 sample, and crop to square areas that are twice the diameter implied by the $R_{\rm circ}$ radii listed in that paper, once converted to the most recent heliocentric distances reported in P20. Some of the selected areas of adjacent cluster cores overlap significantly. The assumed and computed data for these plots are listed in Tables~\ref{P20table}~\&~\ref{G09table}. 

\begin{figure*}%
    \centering
    \includegraphics[width=\textwidth]{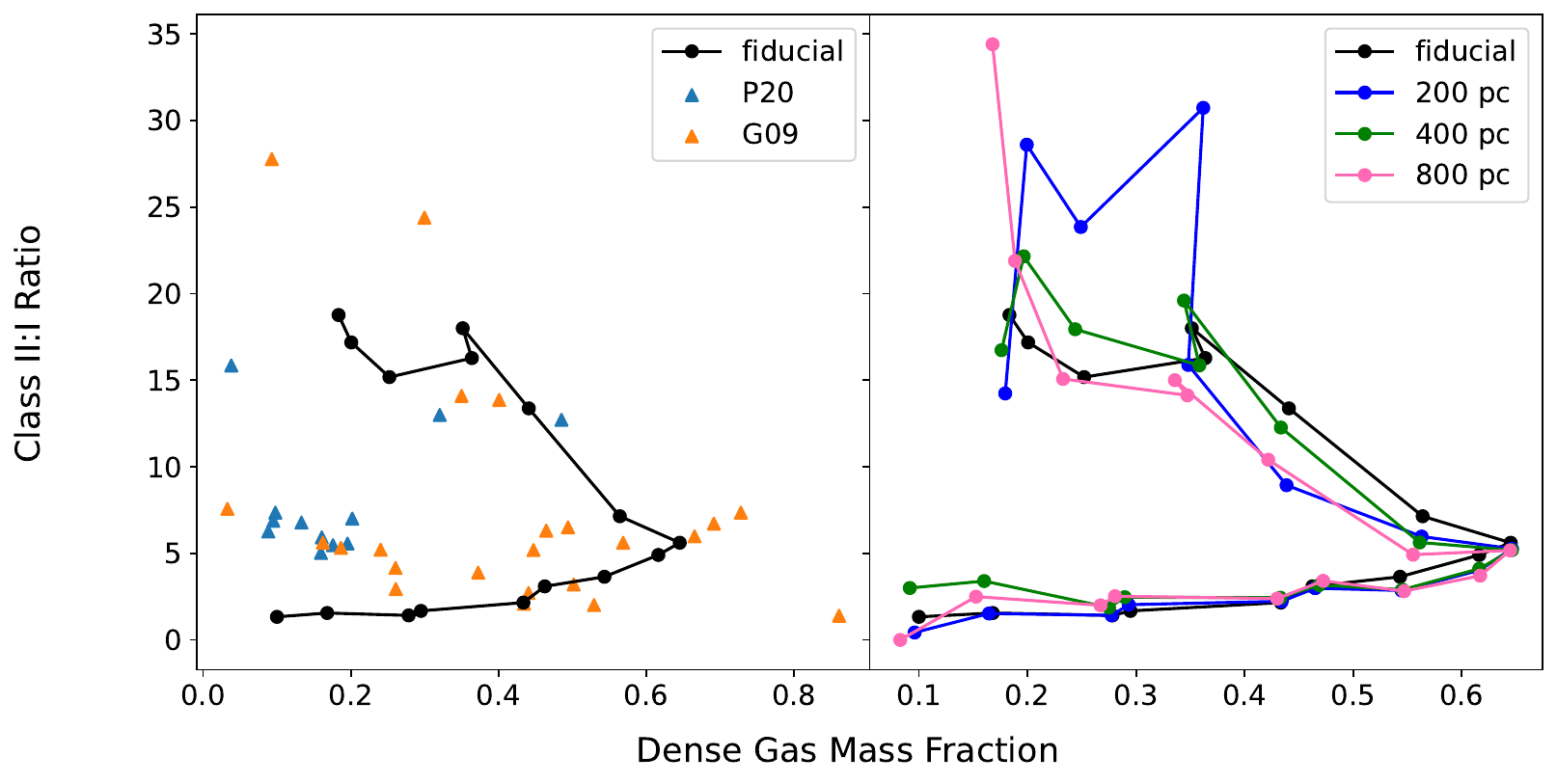}%
    \caption{Dense Gas Mass Fraction versus Class II:I ratio. Left: Blue triangles are values from nearby molecular clouds in P20. Orange triangles represent clusters in those clouds from \citet{Gutea09}. All observed data have been corrected for AGN and edge-on disk contamination, and cropped for coverage consistency and N(H$_{2}) > 10^{21}$~cm$^{-2}$, so they differ to varying degrees from the raw values reported in those works. Black points and line represent the time evolution trajectory for the fiducial analysis of this work, starting from the bottom left. Right: fiducial trajectory overlaid with trajectories from the synthetic analyses at different distances, corrected for AGN contamination. Note that points at high Class II:I ratio are highly uncertain (e.g. Figure \ref{fig:Fig II:I and Disk Fraction}c).}%
    \label{fig:Dense Gas v. Class II:I}%
\end{figure*}

Figure \ref{fig:Dense Gas v. Class II:I}a shows that {\sc starforge} and the clouds in P20 occupy different regions of the Dense Gas Mass Fraction - Class II:I ratio parameter space. The trajectory agrees better with the clusters from \cite{Gutea09}, except for the earliest snapshots. We could correct for this by assuming some amount of ambient star formation occurs in the cloud before the main cluster forms, which would increase the Class II:I ratio, more noticeably in the early and late snapshots that have few Class I and II sources. This supports the implication that {\sc starforge} more closely models the formation of a large cluster rather than star formation in a GMC. Inspection of Tables ~\ref{P20table}~\&~\ref{G09table} indicates that the total gas mass and dense gas mass in the simulation are also more consistent with the ranges reported for the \cite{Gutea09} clusters.

We next apply a correction for AGN to the synthetic observations by removing 4.5 sources per square degree for both Class Is and IIs. We find that the synthetic observation trajectories exhibit strong agreement with each other and the fiducial (Figure \ref{fig:Dense Gas v. Class II:I}b). This is expected, since we add that same density of AGN contaminants at the beginning of the synthetic analysis.

\subsection{Evolution of the Star-Gas Fraction}
\label{subsection:S-G Correlation}
The calculation of the S-G correlation in this work emulates the treatment from P20, allowing us to better compare the outcomes of the two. We calculate the $n$\textsuperscript{th} nearest neighbor distance (NND) for each Class I YSO, for $n=11$ using scipy.spatial.KDTree. KDTree uses the algorithm described by \cite{MM99} to create a binary tree of 3-dimensional nodes (the positions of the sources). This allows for the quick lookup and classification of nearest-neighbors. We use $n=11$ because it is a good compromise of spatial resolution (typically 0.1-2~pc smoothing-scale in nearby clouds) and low relative uncertainty \citep[33\%,][]{CH85}. This choice is consistent with \citet{CH85}, G11, and P20. Using a circular mask with a radius of the NND we calculate the area A of each circular mask, the mean column density in each circle $\Sigma_{\rm C}$, and the ratio $C$ of covered area to total area within each circle. $C$ covers edge effects and is thus almost always unity. From this, we calculate $\Sigma_{\rm gas}$, the gas mass surface density. $\Sigma_{\rm YSO}$ is the measurement of the surface density of YSOs, calculated as

\begin{equation}
    \Sigma_{\rm YSO}=\frac{n-1}{A_{n}C}M_{\rm YSO}
\end{equation}
\citep{CH85}

where $M_{\rm YSO}$ is the adopted mean mass per YSO and $n$, $A_{ n}$ and $C$ are defined above. Except for our fiducial analysis, where we try to avoid as much observational bias as possible, we fix M$_{\rm YSO}$ at 0.5 M\textsubscript{$\odot$} to keep the analysis consistent with P20.

\begin{figure}%
    \centering
    \includegraphics[width=0.47\textwidth]{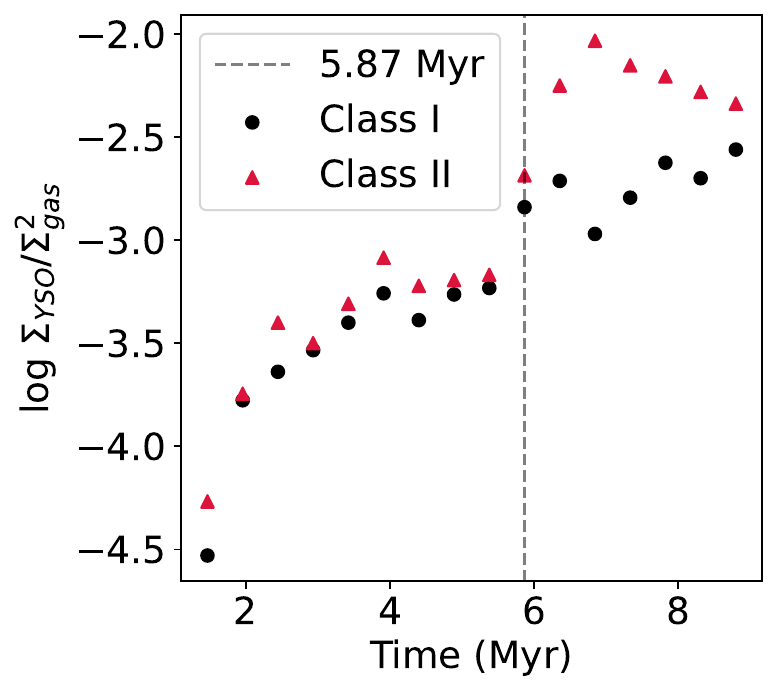}%
    \caption{Median value of $\Sigma_{\rm YSO}/\Sigma_{\rm gas}^{2}$ versus time for both Class I and Class II sources. In addition to showing the increasing stellar density as a function of gas density, these values are closely correlated until $\sim$6~Myr (dashed vertical line). At this time, feedback clearly begins to disrupt the gas (see Figure \ref{fig:Dense Gas Fraction}) thereby inducing decoupling of the gas structure and the YSO distribution.}%
    \label{fig:Sig Star Over Sig Gas Squared}%
\end{figure}

\begin{figure*}%
    \centering
    \includegraphics[width=\textwidth]{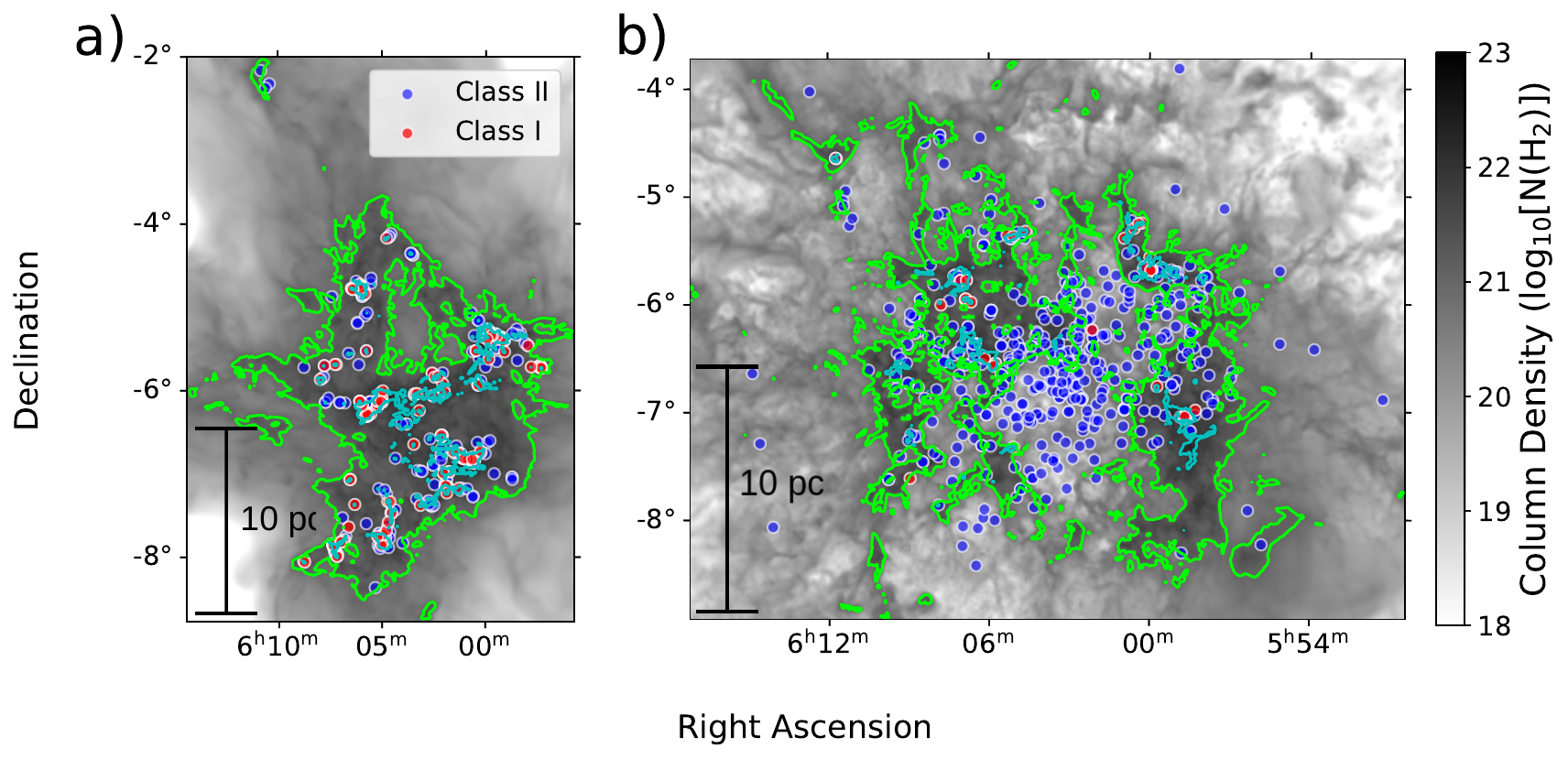}%
    \caption{Projected N(H\textsubscript{2}) column density map with N(H$_{2}) = 10^{21}$~cm$^{-2}$ contour over-plotted in green and N(H$_{2}) = 10^{22}$~cm$^{-2}$ contour in cyan at (a) $\sim$3.4 Myr. (b) $\sim$7.8 Myr. Note in (a) that most YSOs, especially Class I sources, remain close to or within the dense gas cyan contour. In (b) however, many of the YSOs are no longer correlated with the locations of the denser gas at either contour level, indicating YSO-gas decoupling. Existing YSOs (mainly Class IIs and IIIs) remain relatively stationary for the first few million years as the gas dissipates, being bound together by gravity. However, new Class Is continue to form in the denser gas (almost all Class Is are within the cyan contours).}%
    \label{fig:Coupled and UnCoupled}%
\end{figure*}

Figure \ref{fig:Sig Star Over Sig Gas Squared} shows the median $\Sigma_{\rm YSO}/\Sigma_{\rm gas}^{2}$ versus time, which captures the vertical offset and spread around the power-law fit (G11). While anything more than a general positive trend with time showing increasing stellar density as a function of gas density is not immediately clear, the Class I and II values are close to each other for the first $\sim6$~Myr. After this point, the populations no longer appear correlated. This points to a large-scale decoupling of the YSOs from their surrounding gas at around 6-6.5~Myr. This is supported by visual examination of the snapshots. Figure \ref{fig:Coupled and UnCoupled} shows a snapshot before decoupling occurs (3.42 Myr) and a snapshot after decoupling occurs (7.82 Myr). It is clear that nearly all YSOs reside near or within dense gas before the decoupling, but afterwards the two groups are significantly less correlated.

The dense gas mass fraction peaks at $\sim$5.38~Myr (Figure \ref{fig:Dense Gas Fraction}). This is when feedback begins to disperse the cloud (see Figure \ref{fig:Spatial Distribution}), and there is a $\sim$1~Myr lag before the effects are seen in the other statistics. For example, Figure \ref{fig:Fig II:I and Disk Fraction} shows that the number of Class Is drastically declines and the number of Class IIs peaks at $\sim$6.36~Myr. This causes the Class II:I ratio to rise significantly. And, as mentioned above, this is also the time when Class Is and IIs in Figure \ref{fig:Sig Star Over Sig Gas Squared} appear to decouple.

\subsection{Star-Gas Correlation versus Time}

\begin{figure*}
    \centering
    \includegraphics[width=0.75\textwidth]{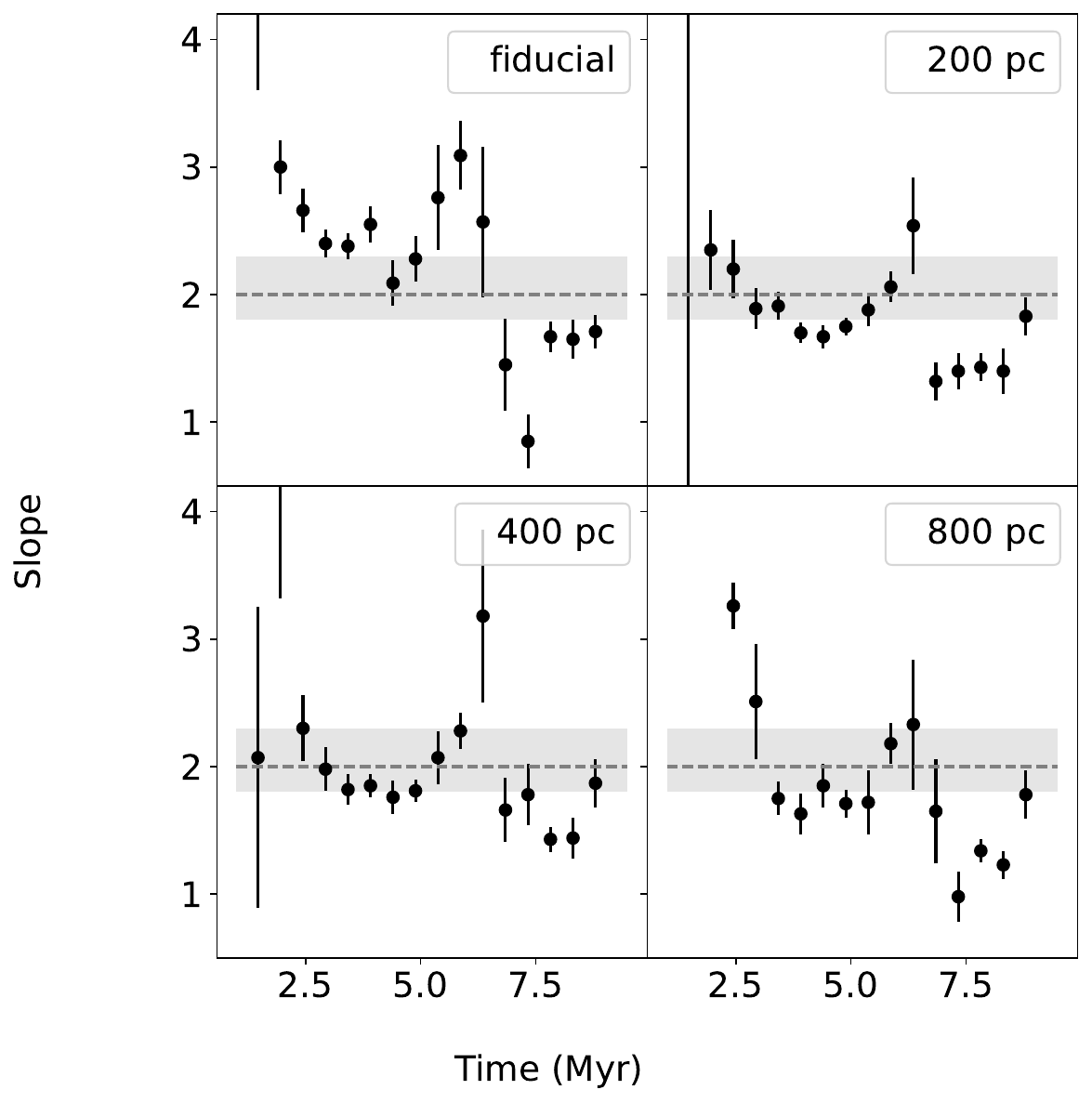}%
    \caption{Slope and uncertainty of the S-G correlation for each snapshot. The shaded region corresponds to the range of values observed by P20. In many of the earlier snapshots, undersampling causes large uncertainties and produces slopes that are discrepant with observations. We limit the $y$ axis to better compare differences between the runs. This obscures the (unreasonable) points of some of the earlier snapshots. See Figures \ref{fig:S-G Default Overview}-\ref{fig:S-G 800 Overview} for full slope and uncertainty values. The leftmost point in the bottom right panel is missing since there is no slope for that individual snapshot (see Figure \ref{fig:S-G 800 Overview}).}%
    \label{fig:S-G With Time Zoom}%
\end{figure*}

Figure \ref{fig:S-G With Time Zoom} shows the slopes and uncertainties of the S-G correlations for the fiducial analysis along with the three sets of synthetic observations as a function of time. Most of the slopes lie relatively close to 2.0, however the well-correlated slopes either lie above or below 2.0, usually localized around 2.4-2.5 or 1.7-1.8. Over half of the fiducial snapshots visually appear to have a tight YSO and gas surface density correlation, with an uncertainty in the slope of $\le0.2$. This provides significant evidence that the power-law relationship for the S-G correlation is a real effect that is a result of underlying physics and not a result of observational bias (see Figure \ref{fig:S-G Default Overview} in the Appendix for the fiducial S-G correlation plots).

However, many of the snapshots are not well-correlated, appearing as a clump of points that lie on the expected line, but do not span a significant range of surface densities. This is especially true for snapshots with fewer than $\sim$100 Class I sources, since this often leads to poorly-constrained slopes with error bars as large as 0.6. This difficulty hinders comparison with previous observations, as many of the observed clouds in G11 and P20 have many more sources and more completely populate the S-G space and thus the S-G correlation. However, the addition of synthetic observation effects, especially adding AGN or removing close neighbors, can artificially compensate for this by filling out the low-density region and depleting the high-density region of the plot, respectively. This is discussed in Section \ref{section:Demonstration of Effects} below. In addition, the S-G correlations for each snapshot of the simulation in the fiducial and 200, 400, and 800 pc synthetic analyses can be found in the Appendix.

Figure \ref{fig:S-G With Time Zoom} illustrates the evolution of the S-G slope as a function of time in the simulation. While the shape, slope, or scatter of the S-G correlation do not change monotonically with time, there are several features that are roughly independent of distance and the presence of the synthetic considerations. This implies that the synthetic observation effects don't obscure the underlying physics, except for in snapshots where low-number (of YSOs) statistics are significant (e.g., the poorly correlated snapshot in Figure \ref{fig:Poor Comparison}). Figure \ref{fig:S-G With Time Zoom} shows the S-G slope declines until around $\sim 4$ Myr (most noticeable at closer distances), at which point the number of Class I sources peaks. From $\sim$4-6 Myr the S-G slope increases as the number of Class II sources continues to rise; the peak in the S-G slope at 6 Myr coincides with the peak in the number of Class II sources. After 6 Myr the S-G slope declines sharply as feedback begins to disperse the cloud in earnest. Many of the snapshots around this time also have poor S-G correlations. Even though much of the cloud gas is dispersed from the central region, the YSOs' dynamics take longer than the gas to respond to the changing gravitational potential. However, star formation still occurs in the remaining pockets of dense gas, maintaining some degree of S-G correlation in the later snapshots (Figure \ref{fig:AGN Example}).

While it is clear that some star-gas statistics evolve with time, the slope of the S-G correlation appears to be relatively constant across the history of the cloud. This is consistent with the observations of G11 and P20 who found little variation in the slopes across a wide collection of GMCs with very different ages. The spread in the S-G slopes and fit uncertainties are significantly larger for the simulations than for the observations of G11 and P20, however. This comparison may not be fully equal, as there is a selection effect on which clouds are actually observed and included for analysis. For this reason the observed clouds may span a narrower range in the cloud lifetime: young clouds with little star formation may not be identified as distinct and/or interesting star-forming regions and thus will be excluded, while older clouds that are in the process of dispersing are excluded since they have little remaining dense material. As a result, the especially-poorly-populated early and late snapshots are not well represented in real data, as it is difficult to find and observe very young and very old clouds. More observational work will need to be done to more effectively compare with these snapshots.

\subsection{Demonstration of Synthetic Effects on the Star-Gas Correlation}
\label{section:Demonstration of Effects}

In this section, we explore how each synthetic effect impacts the apparent S-G correlation. Figures \ref{fig:Comparison} and \ref{fig:Poor Comparison} compare the fiducial S-G correlation with those obtained for five different synthetic effects.

\begin{figure}%
    \centering
    \includegraphics[width=0.47\textwidth]{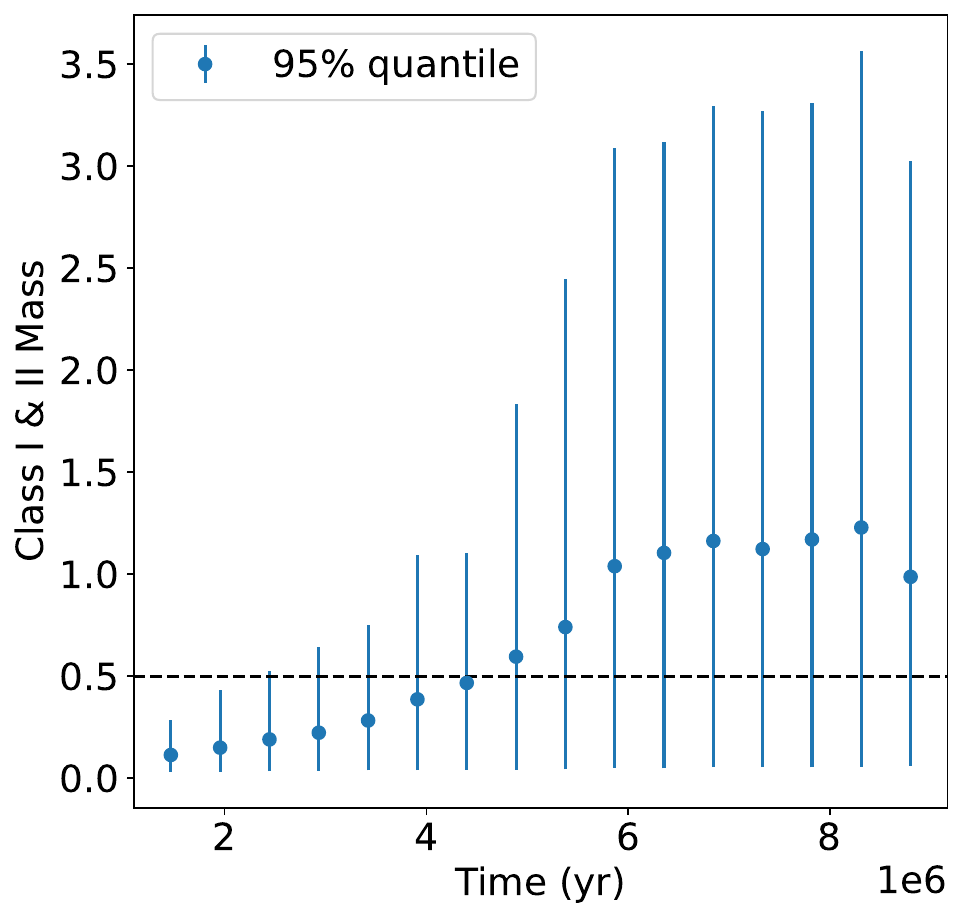}%
    \caption{Average combined Class I and II mass for each snapshot. Error bars represent 95th percentile. It is clear that an assumed mass of 0.5 M$_{\odot}$ does not accurately represent YSO masses at all times. However, this has little effect on the calculation of the S-G correlation (see Section \ref{section:Section 3}).}%
    \label{fig:Avg_mass}%
\end{figure}

\begin{figure*}%
    \centering
    \includegraphics[width=\textwidth]{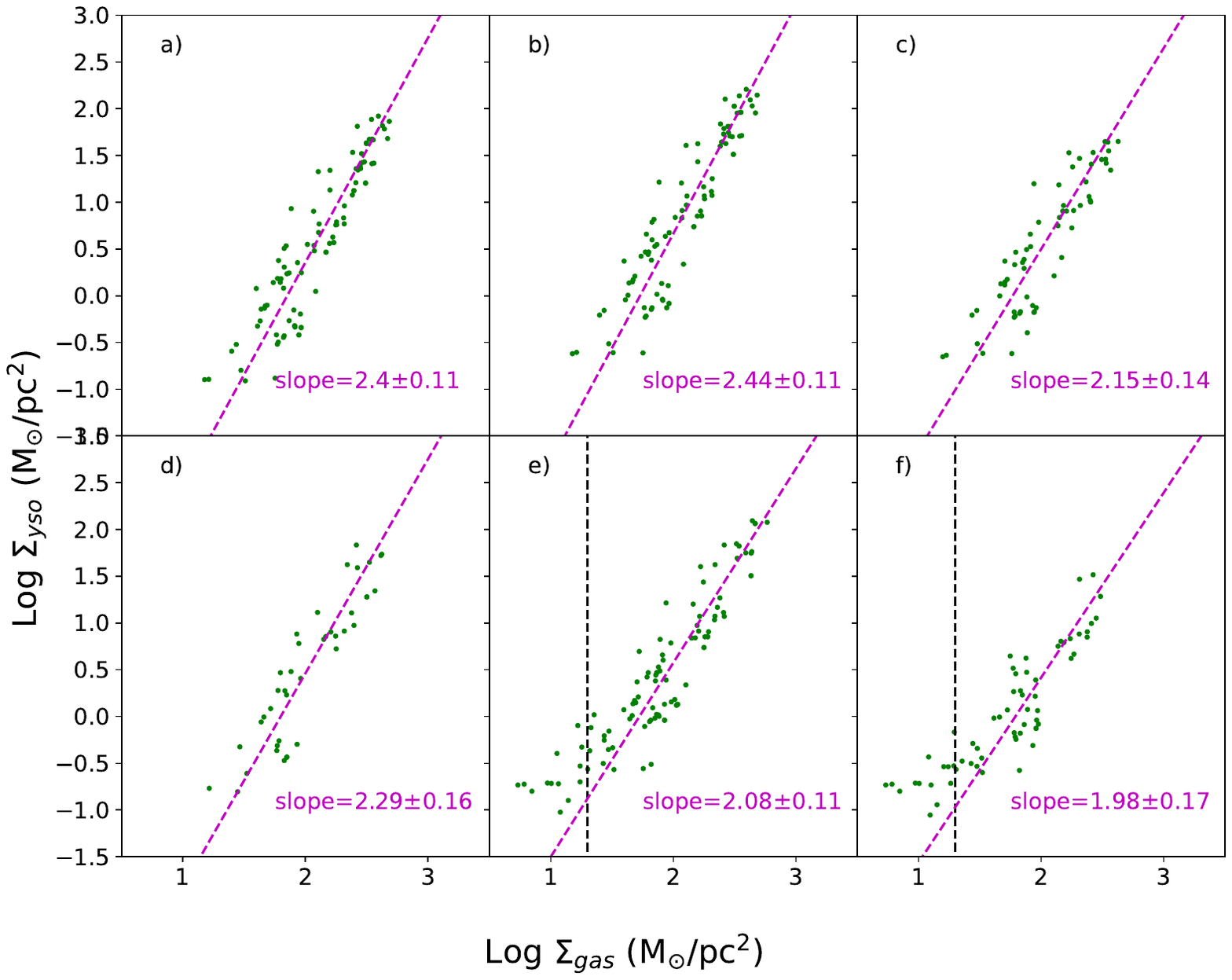}%
    \caption{Comparison of different synthetic observational effects on a well-correlated snapshot. a) The ``fiducial'' analysis with no extra considerations. Each panel b) through e) demonstrates 1 synthetic effect each. b) In the calculation of the S-G correlation, uniform 0.5 M$_{\odot}$ mass for each source is used. There is a slight vertical shift in the points on the plot, but it does not significantly change the slope or the uncertainty of the S-G correlation. We adopt uniform mass for the rest of the panels. c) The removal of close neighbors that would have been indistinguishable by \textit{Spitzer}. This predominantly removes high-density points, lowering the slope. d) The removal of low-mass sources that would have been undetected by \textit{Spitzer}. This removes points without visible bias towards density, increasing the uncertainty in the slope. e) The addition of AGN. This predominantly adds low-density sources, lowering the slope. f) All previous synthetic effects at once. The slope is much closer to 2. Black dashed lines represent a density cutoff imposed to account for the presence of AGN. The slopes and best-fit lines for e) and f) are only based on points to the right of the black line.}%
    \label{fig:Comparison}%
\end{figure*}

\begin{figure*}%
    \centering
    \includegraphics[width=\textwidth]{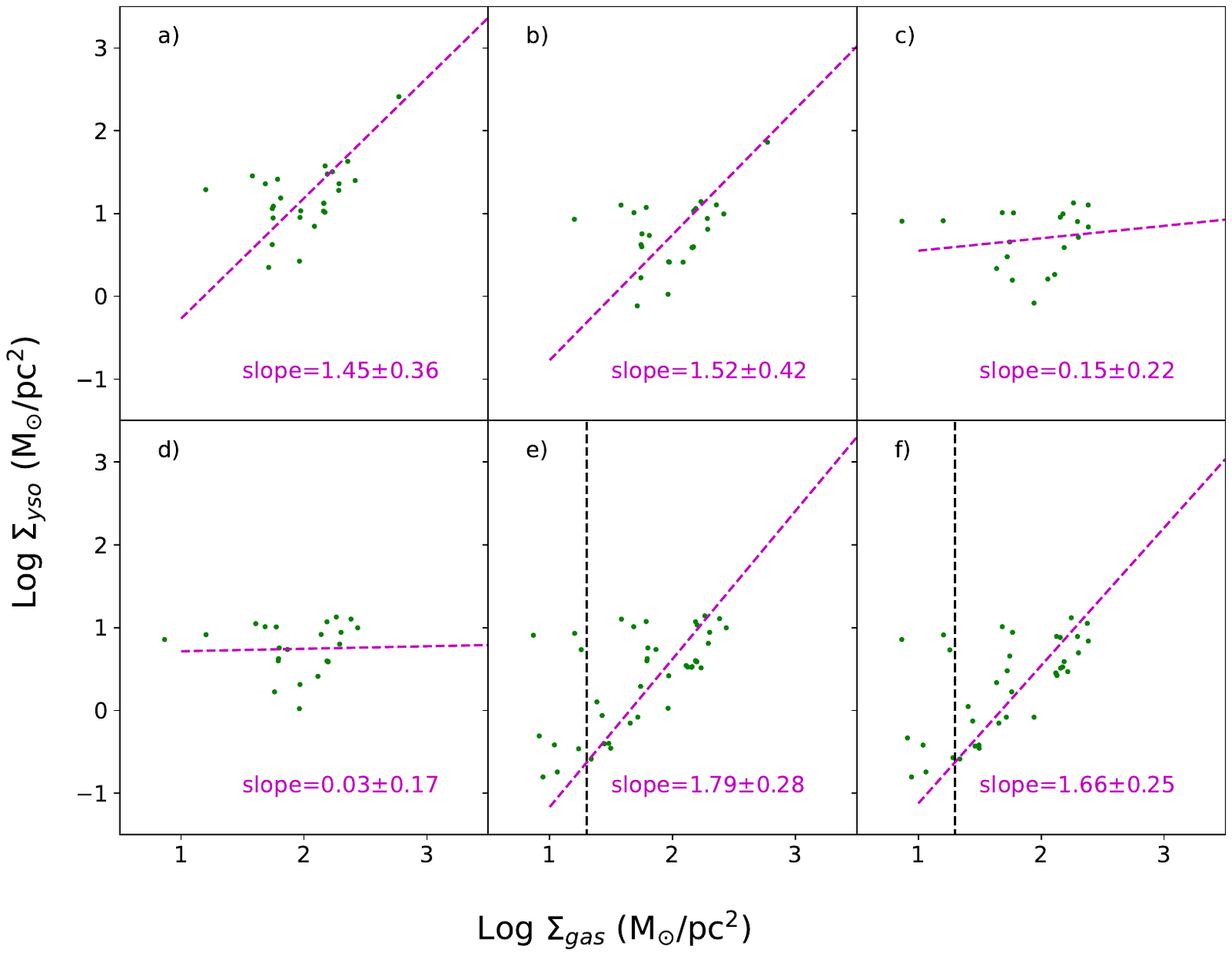}%
    \caption{Comparison of different synthetic observational effects on a poorly-correlated snapshot. Features of figure the same as in Figure \ref{fig:Comparison}. Note that this snapshot is particularly sensitive to the removal of a single high-density point.}%
    \label{fig:Poor Comparison}%
\end{figure*}

The first effect we add to the synthetic observation is the adoption of a uniform YSO mass. Figure \ref{fig:Avg_mass} shows the mean and spread of  YSO masses in the simulation, and as can clearly be seen, adopting a fixed average mass does not well-represent the true average mass, which varies by a factor of $\sim 10$ over time. However, since individual real YSO masses cannot be directly measured, observational analyses such as P20 must adopt some approximation. Figures \ref{fig:Comparison}b and \ref{fig:Poor Comparison}b illustrate the S-G correlation assuming a uniform YSO mass of 0.5 M$_{\odot}$. Comparing panels (a) and (b) indicates that using the true mass of the sources has little effect on the S-G correlation. While the points move slightly vertically, the slopes change by less than 0.1. Consequently, source mass appears to have a relatively minor impact on the S-G correlation. Considering the minor effects, the uniform mass is used in the rest of the demonstration.

The impact of the removal of close neighbors is more significant. When multiple close sources appear as a single source, the effect is to remove many of the highest density points in the S-G relationship shown in Figure 9c. This, in turn, has a flattening effect on the power-law slope, for example, bringing the slopes of most snapshots (all snapshots between 2 and 6.5 Myr) within 2$\sigma$ of 2.0 (see Figure \ref{fig:S-G With Time Zoom}). The earlier and later snapshots tend to be (often significantly) less well-sampled, which likely explains their inconsistent slopes (as shown in Figure \ref{fig:Poor Comparison}). The impact of this effect increases with distance, as the 5\arcsec \,minimum separation imposed on the YSO lists translates to larger physical separations. This can be observed qualitatively in Figures \ref{fig:S-G Default Overview} - \ref{fig:S-G 800 Overview} in the Appendix.

Figure \ref{fig:Comparison}d illustrates the effect of detection limits on the S-G correlation. We find that implementing a mass sensitivity limit significantly decreases the number of sources at all densities, which increases the fit uncertainties across all snapshots. However, this does not significantly change the slope in well-correlated snapshots. In contrast, Figure \ref{fig:Poor Comparison} shows that the addition or removal of a single point can significantly change the slope in a snapshot with fewer YSOs. This effect is also more extreme at larger distances. Specifically, at 200 and 400 pc, the mass limit is the Hydrogen-burning limit: 0.08 M$_{\odot}$, while the limit at 800 pc is 0.2 M$_{\odot}$, significantly reducing the number of sources with which to calculate the S-G correlation. Compare Figure \ref{fig:S-G 800 Overview} with Figures \ref{fig:S-G 200 Overview} and \ref{fig:S-G 400 Overview} in the Appendix for a visualization of how the number of sources decreases with increasing distance.

Next we investigate the impact of AGN contamination on the S-G correlation. The addition of AGN has a significant impact on the S-G correlation as shown in Figures \ref{fig:Comparison}e and \ref{fig:Poor Comparison}e. Since the AGN are uniformly randomly distributed throughout the field of view, they add a relatively constant ($\Sigma_{\rm YSO}=\:\sim$0.3, $\sim$0.1, $\sim$0.03 M$_{\odot}$/pc$^{2}$ at 200, 400, and 800 pc, respectively) ``foot'' of points to the bottom of the S-G correlation. This disproportionately affects low $\Sigma_{\rm YSO}$ regions and artificially flattens the power law. The flattening increases with distance, so much so that the slope of the S-G correlation for every snapshot at 200 pc and 400 pc would lie below 2 at all times. However, we follow observational convention and implement a column density cutoff when fitting the slope as described below.

For nearby clouds, the number of YSOs observed at relatively low column densities is small, which causes observations of those regions to be dominated by AGN contaminants. To deal with the similar issue of our synthetic AGN, we adopt the same approach as G11 and remove YSOs in our catalog in regions with ${\rm log}(\Sigma_{\rm gas})$ $<$ 1.3 M$_{\odot}$ pc$^{-2}$ and refit the remainder. This is demonstrated in Figures \ref{fig:Comparison}e,f and \ref{fig:Poor Comparison}e,f. Applying this treatment to the synthetic observations with AGN confirms that such a cut is justified to minimize the bias of the fit caused by AGN contamination. After applying this cut, most slopes steepen and approach the expected value of $\sim$2.0 (Figure \ref{fig:S-G With Time Zoom}).

\section{DISCUSSION}
\label{section:Section 4}

\subsection{Implications for the S-G Correlation}
The presence of many well-correlated S-G relationships for the fiducial {\sc starforge} run implies that the S-G correlation is a physical phenomenon and not solely the result of observational biases. However, the addition of synthetic observation considerations does artificially lower the slope of the S-G correlation for many of the snapshots, generally increasing agreement with observations. This begs the question of whether the very consistent value of 2 determined by P20 is partially caused by observational effects. In this case, the S-G correlation slope is not as invariant and universal as it appears in P20. However, in the latest snapshots, which have better agreement in Class II:I ratio versus dense gas mass fraction with P20 clouds (Figure \ref{fig:Dense Gas v. Class II:I}), the S-G correlation slopes are much lower than observed.

Nonetheless, it is striking that the broad range of evolutionary stages spanned by one cloud, modeled with all key physical effects, produce a relatively uniform power-law slope. Once star formation is underway, stellar feedback helps to regulate the relationship between dense gas and YSOs. Clouds with particularly high Class II to I ratios are likely dominated by stellar feedback and in the process of cloud dispersal. Follow-up observations that minimize observational effects are required to fully constrain if and the extent to which these biases conspire to produce an S-G power-law slope of $\sim 2$.

\subsection{Comparison to Previous Work}

\cite{Cheea21} find that GMCs in nine nearby disc galaxies usually disperse within $\sim3$ Myr after unembedded high-mass stars emerge. While not directly measured in this work, we believe the first high-mass stars likely emerge shortly before feedback begins dispersing the cloud in earnest. We estimate dispersal to become qualitatively significant sometime between $\sim5.4-6.4$ Myr, as described in Section 3. And, considering the GMC is nearly completely dispersed by 8.8 Myr, the simulations are consistent with the observed $\lesssim3$ Myr time frame, as well as the $\sim10$ Myr total cloud lifetime they estimate.

As mentioned in the Introduction, P20 adapted HD simulations by \cite{Qiaea15} to create synthetic observations of their 12 observed clouds. These synthetic observations included 2D projection and neighbor removal. The HD simulations produced slopes between $2.3-2.7$, higher than the observed $1.8-2.3$. The simulations are also limited in density dynamic range compared to some of the clouds they model (see Figure 6 in P20). However, the simulated slopes are similar to the values of $2.0-3.$0 we observe in the fiducial run (before cloud dispersal and excluding the first two snapshots, see \ref{tab:Table 2}). Caution is required when comparing with these simulations since P20 modeled 12 different clouds at a single time, while this work models one cloud at many different times. Regardless, the main improvement of {\sc starforge} over the simulations in \cite{Qiaea15} is more realistic physics, especially magnetic fields and kinematic feedback. While magnetic fields do not play a very significant role in setting the slope of the S-G correlation, kinematic feedback allows {\sc starforge} to evolve the GMC without driven turbulence \citep[which was necessary for the simulations in][]{Qiaea15}. While the STARFORGE simulation starts with an initial turbulent setup, the turbulence, evolution, and dispersal of the cloud are regulated entirely by stellar feedback.

\subsection{Model and Analysis Caveats}

While {\sc starforge} faithfully reproduces the S-G correlation in many snapshots, there are some areas for improvement. For example, Figure \ref{fig:Dense Gas v. Class II:I}, which displays the Class II to I ratio versus dense gas mass fraction, shows that the simulation exhibits poor agreement with the P20 clouds. In contrast, the simulation agrees better with the cluster data from \cite{Gutea09}. This implies that the {\sc starforge} simulation analyzed here produces something that is closer to a smaller denser structure (i.e., a cluster) than the stellar complexes formed in the GMCs of the P20 sample, which may be characterized by a longer and richer star formation history.

The drastic evolution in the number of Class Is with time highlights that the simulation SFR is not constant, in contrast with the assumption of constant SFR made by G11, P20, and others when using class ratios (to infer age, for example). Figure \ref{fig:Fig II:I and Disk Fraction} shows that this leads to Class II:I ratios that vary much more than in the P20 observations. \citet{MGK22} argued that a variable SFR similar to that produced by the {\sc starforge} simulation is necessary to explain the ensemble of Class II:I ratios and disk fractions in nearby clusters. The agreement between this model and the {\sc starforge} data (Figure \ref{fig:Model Comparison}) provides more evidence that {\sc starforge} produces something more similar to a monolithic cluster than a full GMC (i.e., with several smaller, distinct clusters). 

However, we caution that here we only analyze one simulation that aims to model the typical conditions of a Milky Way cloud. Future work is needed to explore the broad range of conditions modeled in the {\sc starforge} simulation suite, which includes clouds with varying initial magnetic field, turbulence, interstellar radiation field, surface density, cloud size, and cloud initialization \citep{Gusea22}. In particular, the initial cloud setup, that of a uniform density sphere, is a significant oversimplification of the complexity of forming and accreting molecular clouds. Overall, agreement with both data sets would likely be improved by using more realistic initial conditions. For example, a slower star formation start could increase the Class II:I ratio in the early snapshots, improving agreement. Simulations that begin with more realistic cloud initialization, such as a driven-turbulence sphere \citep{LaneGrudic2022} or models that follow cloud formation from galaxy scales \citep[][Hopkins et al. 2023 in prep.]{HuKrumholz2023,Ganguly2023}, are likely necessary to advance agreement between the {\sc starforge} framework and observations.

One recent interesting aspect of {\sc starforge} comes from \cite{Gruea23}, who ran 100 2000 M$_{\odot}$ STARFORGE simulations and found a sharp mass cutoff on the IMF at 28 M$_{\odot}$. This is in contrast to a simulation with similar parameters but 10 times the mass, which generated a 44 M$_{\odot}$ star, and a simulation with 10 times the gas surface density which generated a 107 M$_{\odot}$ star. They suggest that the STARFORGE IMF has a high-mass cutoff that depends on the environment. This cutoff is generally different from the canonical 100$-$150M$_{\odot}$ cutoff, which they conclude implies that the IMF cannot be reproduced in small clouds simply by randomly sampling from the full IMF.

Here we also outline some inconsistencies in our data processing. The bounds on the cropped field of view (see Figure \ref{fig:AGN Example}) are set by the furthest extent of the N(H$_{2}) = 10^{21}$~cm$^{-2}$ contour. This occasionally causes larger-than-intended fields of view when an area of gas denser than N(H$_{2}) = 10^{21}$~cm$^{-2}$ is present away from the central cluster. The only major impact this has on the analysis is to increase the number of AGN when calculating the S-G correlation. However, this impact is largely mitigated by the low density cut discussed previously. We also neglect a number of steps that would be needed to complete a fully ``apples-to-apples" comparison with the observations. For example, we do not use radiative post-processing to construct the YSO SEDs \citep[e.g.,][]{Offner2012}. Nor do we construct synthetic dust continuum maps in order to compute the column density \citep[e.g.,][]{Juvela2019}. These steps would allow us to apply the observational biases, such as the detection limits, more directly. However, we expect any impact on the S-G slope to be minimal, since the YSO positions and relative amounts of dense gas would be unchanged.

\section{CONCLUSIONS}
\label{section:Section 5}

In this study, we examine a 20,000 M$_{\odot}$ star-forming cloud in the {\sc starforge} simulation suite in order to investigate the presence and evolution of the S-G correlation. To effectively do so, we create synthetic observations to compare with previous observational work, specifically P20 and G11. These synthetic observations include 2D projection of gas and star particle distributions at multiple distances, an age-to-Class conversion for the simulated stars using an exponential decay model, AGN contamination, a low stellar mass cutoff, the removal of close (unresolved) neighbors, gas map smoothing to mimic limited angular resolution, and a field-of-view crop at a gas column density of N(H$_{2}) = 10^{21}$~cm$^{-2}$.

Since most of these effects depend on distance, we place each cloud at 200, 400 and 800 pc to mimic the distances of star-forming regions observed by G11 and P20. This changes the angular size of the cloud, the number of AGN, the mass sensitivity limit, and the neighbor threshold. From these synthetic observations, we examine the dense gas fraction, YSO distribution and frequency, and the S-G correlation for the fiducial analysis and for the synthetic analyses at each distance.

We find that the {\sc starforge} simulation successfully reproduces the S-G correlation in many snapshots and exhibits a typical S-G slope within $1\sigma$ of the observed slope of 2. The presence of the S-G correlation both with and without accounting for observational effects implies that this is a real relationship that is a product of the underlying physical processes. However, observational biases, such as AGN contamination, appear to strengthen the S-G correlation, reduce time variation and promote a slope closer to 2. 

We find that the Class II:I ratios and dense gas fraction characteristic of the {\sc starforge} simulation exhibit better agreement with those of the clusters in the \cite{Gutea09} sample than the stellar complexes forming in the clouds in P20. No regions in either observational study match the low Class II:I ratios found at early times ($<3$ Myr) in the simulation. This implies that the P20 and \cite{Gutea09} clouds/clusters form stars at a low rate for a few million years. Thus, bias in cloud selection, which favors actively star-forming clouds with significant amounts of dense gas, possibly also contributes to the apparent universality of the S-G correlation.

The present study only considers the S-G correlation under one set of typical simulated cloud conditions. Future work is needed to examine the impact of cloud properties and more realistic initial conditions on the S-G correlation.

\section*{Acknowledgements}

SSRO and RG acknowledge funding support for this work from NSF AAG grants 2107340 and 2107705. SSRO acknowledges support by NSF through CAREER award 1748571, AST-2107340 and AST-2107942, by NASA through grants 80NSSC20K0507 and 80NSSC23K0476, and by the Oden Institute through a Moncrief Grand Challenge award. Support for MYG was provided by NASA through the NASA Hubble Fellowship grant \#HST-HF2-51479 awarded by the Space Telescope Science Institute, which is operated by the Association of Universities for Research in Astronomy, Inc., for NASA, under contract NAS5-26555. The simulation was run on the Frontera supercomputer with LRAC award AST21002. This research is part of the Frontera computing project at the Texas Advanced Computing Center. Frontera is made possible by National Science Foundation award OAC-1818253.

SM acknowledges the Massachusetts Space Grant Consortium in their support of his participation in the Summer 2022 Five-College Astronomy Undergraduate Internship Program.

This work relies on products from SESNA, the Spitzer Extended Solar Neighborhood Archive, and associated Herschel data, developed with generous support from NASA ADAP grants NNX11AD14G, NNX13AF08G, NNX15AF05G, and NNX17AF24G, and NASA JPL/Caltech Herschel support grant 1489384.

SM and RG acknowledge and thank Ronald Snell and S.T. Megeath for their support in constructing the paper.

We would also like to thank the referee for their helpful comments.

\software{Astropy \citep{Astropy}, h5py (\url{http://www.h5py.org/}), Matplotlib \citep{Matplotlib}, NumPy \citep{Numpy}, SciPy \citep{Scipy}, yt \citep{yt}}



\bibliographystyle{aasjournal}
\bibliography{STARFORGE}




\appendix
\section{Tables}
In Tables \ref{tab:Table 1} and \ref{tab:Table 2}, we present various statistics extracted from the analysis of the fiducial snapshot. We present statistics of the clouds and cluster cores from P20 and G09 used in our analysis in Tables \ref{P20table} and \ref{G09table}. We have updated the data from P20 and G09 to the latest datasets from SESNA and \textit{Herschel}, corrected for AGN and edge-on disk contamination, and adopted distances from P20.
\begin{deluxetable}{|c|c|c|c|c|c|c|c|c|}[h]
\centering
\tablecolumns{9}
\tablewidth{\textwidth}
\tablehead{Time & M$_{\rm gas}$ & M$_{\rm dense}$ & M$_{\rm dense}$/M$_{\rm gas}$ & N$_{\rm Class I}$ & N$_{\rm Class II}$ & Mean Age (All) & Mean Age Class I & Mean Age Class I \& II}
\startdata
(Myr) & ($10^{3}$ M$_{\odot}$) & ($10^{3}$ M$_{\odot}$) &      &     &     & (Myr) & (Myr) & (Myr)                   \\ \hline
1.47  & 2.78                   & 0.278                  & 0.10 & 9   & 12  & 0.23  & 0.23  & 0.24                    \\ \hline
1.96  & 4.03                   & 0.676                  & 0.17 & 27  & 42  & 0.39  & 0.18  & 0.36                    \\ \hline
2.44  & 5.40                   & 1.50                   & 0.28 & 60  & 85  & 0.51  & 0.21  & 0.46                    \\ \hline
2.93  & 6.12                   & 1.80                   & 0.30 & 99  & 166 & 0.64  & 0.23  & 0.56                    \\ \hline
3.42  & 7.58                   & 3.29                   & 0.43 & 134 & 289 & 0.79  & 0.23  & 0.62                    \\ \hline
3.91  & 7.65                   & 3.54                   & 0.46 & 127 & 393 & 1.05  & 0.30  & 0.81                    \\ \hline
4.40  & 8.90                   & 4.84                   & 0.54 & 122 & 445 & 1.31  & 0.33  & 0.98                    \\ \hline
4.89  & 9.65                   & 5.95                   & 0.62 & 108 & 530 & 1.56  & 0.38  & 1.12                    \\ \hline
5.38  & 9.84                   & 6.35                   & 0.65 & 98  & 551 & 1.86  & 0.31  & 1.35                    \\ \hline
5.87  & 8.25                   & 4.65                   & 0.56 & 84  & 601 & 2.14  & 0.32  & 1.49                    \\ \hline
6.36  & 6.32                   & 2.78                   & 0.44 & 43  & 576 & 2.55  & 0.54  & 1.81                    \\ \hline
6.84  & 4.82                   & 1.69                   & 0.35 & 30  & 541 & 2.94  & 0.49  & 2.14                    \\ \hline
7.33  & 4.01                   & 1.46                   & 0.36 & 29  & 475 & 3.34  & 0.36  & 2.40                    \\ \hline
7.82  & 2.74                   & 0.691                  & 0.25 & 28  & 429 & 3.74  & 0.40  & 2.61                    \\ \hline
8.31  & 2.07                   & 0.415                  & 0.20 & 22  & 383 & 4.18  & 0.38  & 2.92                    \\ \hline
8.80  & 1.96                   & 0.360                  & 0.18 & 17  & 323 & 4.61  & 0.29  & 3.15                    \\ \hline%
\enddata
\caption{Table of various fiducial snapshot statistics.}
\label{tab:Table 1}
\end{deluxetable}

\begin{deluxetable}{|c|c|c|c|c|c|c|}
\centering
\tablecolumns{10}
\tablehead{Time & Mean Mass Class I \& II & Class II:I & S-G Slope & Med. $\Sigma_{\rm Class I}/\Sigma_{\rm gas}^{2}$ & Med. $\Sigma_{\rm Class II}/\Sigma_{\rm gas}^{2}$ & SFE\tablenotemark{a}}
\startdata
(Myr)& (M$_{\odot}$) &            &           & (pc$^{2}$/M$_{\odot}$) & (pc$^{2}$/M$_{\odot}$) &                        \\ \hline
1.47 & 0.11          & 1.33       & 7.60      & -3.91                  & -3.68                  & 0.38\%                 \\ \hline
1.96 & 0.15          & 1.56       & 3.03      & -3.21                  & -3.21                  & 0.85\%                 \\ \hline
2.44 & 0.19          & 1.42       & 2.60      & -3.31                  & -3.07                  & 1.32\%                 \\ \hline
2.93 & 0.22          & 1.68       & 2.44      & -3.22                  & -3.20                  & 2.12\%                 \\ \hline
3.42 & 0.28          & 2.16       & 2.40      & -3.25                  & -3.18                  & 2.71\%                 \\ \hline
3.91 & 0.39          & 3.09       & 2.48      & -3.31                  & -3.21                  & 3.29\%                 \\ \hline
4.40 & 0.47          & 3.65       & 2.00      & -3.49                  & -3.34                  & 3.09\%                 \\ \hline
4.89 & 0.60          & 4.91       & 2.30      & -3.49                  & -3.45                  & 3.20\%                 \\ \hline
5.38 & 0.74          & 5.62       & 2.76      & -3.44                  & -3.37                  & 3.19\%                 \\ \hline
5.87 & 1.04          & 7.15       & 3.12      & -3.26                  & -3.11                  & 3.99\%                 \\ \hline
6.36 & 1.10          & 13.4       & 2.35      & -3.09                  & -2.67                  & 4.67\%                 \\ \hline
6.84 & 1.16          & 18.0       & 1.52      & -3.53                  & -2.43                  & 5.59\%                 \\ \hline
7.33 & 1.12          & 16.4       & 0.89      & -3.21                  & -2.58                  & 5.91\%                 \\ \hline
7.82 & 1.17          & 15.3       & 1.65      & -3.05                  & -2.64                  & 7.70\%                 \\ \hline
8.31 & 1.23          & 17.4       & 1.61      & -3.10                  & -2.71                  & 8.91\%                 \\ \hline
8.80 & 0.99          & 19.0       & 1.71      & -2.87                  & -2.64                  & 7.98\%                 \\ \hline%
\enddata
\tablenotetext{a}{Calculated using M$_{\rm gas}$ and Class I and II YSOs, assuming constant 0.5 M$_{\odot}$ YSO mass.}
\caption{Table of various fiducial snapshot statistics.}
\label{tab:Table 2}
\end{deluxetable}

\begin{deluxetable}{cccccccccc}
\tabletypesize{\scriptsize}
\rotate
\tablecaption{P20 Cloud Statistics\label{P20table}}
\tablewidth{0pt}
\tablehead{\colhead{Name} & \colhead{Dist. (pc)} & \colhead{$M_{\rm gas}$} & \colhead{$M_{dense}$} & \colhead{Class II} & \colhead{Class I} & \colhead{Area (pc$^2$)} & \colhead{$M_{dense}/M_{\rm gas}$} & \colhead{C2/C1} & \colhead{SFE}}
\startdata
Ophiuchus & 140 & 2200 & 390 & 256 & 46 & 31.0 & 0.18 & 5.50 & 0.06 \\
Perseus & 290 & 5300 & 840 & 346 & 68 & 85.0 & 0.16 & 5.00 & 0.04 \\
Orion A & 415 & 49000 & 24000 & 2013 & 158 & 350. & 0.49 & 13.0 & 0.02 \\
Orion B & 415 & 10000 & 2000 & 413 & 58 & 130. & 0.20 & 7.00 & 0.02 \\
Aquila N & 430 & 9900 & 940 & 331 & 48 & 110. & 0.10 & 6.90 & 0.02 \\
Aquila S & 430 & 37000 & 6000 & 734 & 124 & 290. & 0.16 & 5.90 & 0.01 \\
Mon OB1 & 800 & 21000 & 4000 & 445 & 80 & 270. & 0.20 & 5.60 & 0.01 \\
S140 & 750 & 5400 & 710 & 256 & 37 & 80.0 & 0.13 & 6.80 & 0.03 \\
AFGL 490 & 1000 & 7000 & 690 & 266 & 36 & 95.0 & 0.10 & 7.40 & 0.02 \\
Cepheus OB3 & 800 & 73000 & 2800 & 1989 & 125 & 950. & 0.04 & 16.0 & 0.01 \\
Mon R2 & 860 & 35000 & 3100 & 718 & 114 & 810. & 0.09 & 6.30 & 0.01 \\
Cygnus X & 1400 & 2100000 & 680000 & 16899 & 1300 & 1500 & 0.32 & 13.0 & 0.03\tablenotemark{a} \\
\enddata
\tablenotetext{a}{Stellar completeness corrected by 0.163 following P20.}
\end{deluxetable}

\begin{deluxetable}{cccccccccc}
\tabletypesize{\scriptsize}
\rotate
\tablecaption{G09 Cluster Core Statistics\label{G09table}}
\tablewidth{0pt}
\tablehead{\colhead{Name} & \colhead{Dist. (pc)} & \colhead{$M_{\rm gas}$} & \colhead{$M_{\rm dense}$} & \colhead{Class II} & \colhead{Class I} & \colhead{Area (pc$^2$)} & \colhead{$M_{\rm dense}/M_{\rm gas}$} & \colhead{C2/C1} & \colhead{SFE}}
\startdata
Oph L1688 Core-1  & 140 & 660 & 290 & 164 & 31 & 4.70 & 0.45 & 5.20 & 0.13 \\
Oph L1688 Core-2  & 140 & 300 & 200 & 88 & 14 & 1.40 & 0.67 & 6.00 & 0.14 \\
IC348 Core-1  & 290 & 320 & 96 & 135 & 5 & 4.10 & 0.30 & 24.0 & 0.18 \\
IC348 Core-2  & 290 & 220 & 88 & 62 & 4 & 1.60 & 0.40 & 14.0 & 0.13 \\
IC348 Core-3  & 290 & 230 & 79 & 81 & 5 & 2.00 & 0.35 & 14.0 & 0.16 \\
NGC 1333 Core  & 290 & 700 & 310 & 93 & 34 & 6.20 & 0.44 & 2.70 & 0.08 \\
Serpens Core  & 430 & 810 & 430 & 57 & 28 & 4.80 & 0.53 & 2.00 & 0.05 \\
MWC 297 Core  & 430 & 390 & 94 & 16 & 3 & 2.20 & 0.24 & 5.20 & 0.02 \\
S140 Core  & 750 & 1000 & 440 & 26 & 12 & 7.80 & 0.43 & 2.10 & 0.02 \\
AFGL490 Core-1  & 1000 & 4300 & 700 & 217 & 38 & 49.0 & 0.16 & 5.60 & 0.03 \\
AFGL490 Core-2  & 1000 & 1200 & 320 & 68 & 23 & 12.0 & 0.26 & 2.90 & 0.04 \\
Cep C Core-1  & 800 & 1600 & 790 & 85 & 26 & 8.80 & 0.50 & 3.20 & 0.03 \\
Cep C Core-2  & 800 & 350 & 300 & 11 & 8 & 1.20 & 0.86 & 1.40 & 0.03 \\
Cep A Core-1  & 800 & 1700 & 780 & 86 & 13 & 8.90 & 0.46 & 6.30 & 0.03 \\
Cep A Core-2  & 800 & 650 & 470 & 28 & 3 & 2.10 & 0.73 & 7.40 & 0.02 \\
L1211 Core-1  & 800 & 1300 & 330 & 71 & 17 & 10.0 & 0.26 & 4.20 & 0.03 \\
L1211 Core-2  & 800 & 1500 & 280 & 74 & 14 & 15.0 & 0.19 & 5.30 & 0.03 \\
Mon R2 Core  & 860 & 2800 & 1400 & 220 & 33 & 18.0 & 0.49 & 6.50 & 0.04 \\
GGD 12-15 Core  & 860 & 1200 & 660 & 97 & 17 & 7.20 & 0.57 & 5.60 & 0.05 \\
GGD 17 Core  & 860 & 560 & 210 & 33 & 8 & 4.70 & 0.37 & 3.90 & 0.04 \\
IRAS 06046-0603 Core  & 860 & 650 & 21 & 22 & 3 & 9.40 & 0.03 & 7.60 & 0.02 \\
S106 Core  & 1400 & 7400 & 5100 & 85 & 12 & 31.0 & 0.69 & 6.70 & 0.04\tablenotemark{a} \\
BD+40$^{\circ}$4124 Core  & 1400 & 18000 & 1700 & 222 & 8 & 160. & 0.09 & 28.0 & 0.04\tablenotemark{a} \\
\enddata
\tablenotetext{a}{Stellar completeness corrected by 0.163 following P20.}
\end{deluxetable}

\clearpage

\section{S-G Correlation Plots}

\begin{figure*}[h]%
    \centering
    \includegraphics[width=.9\textwidth]{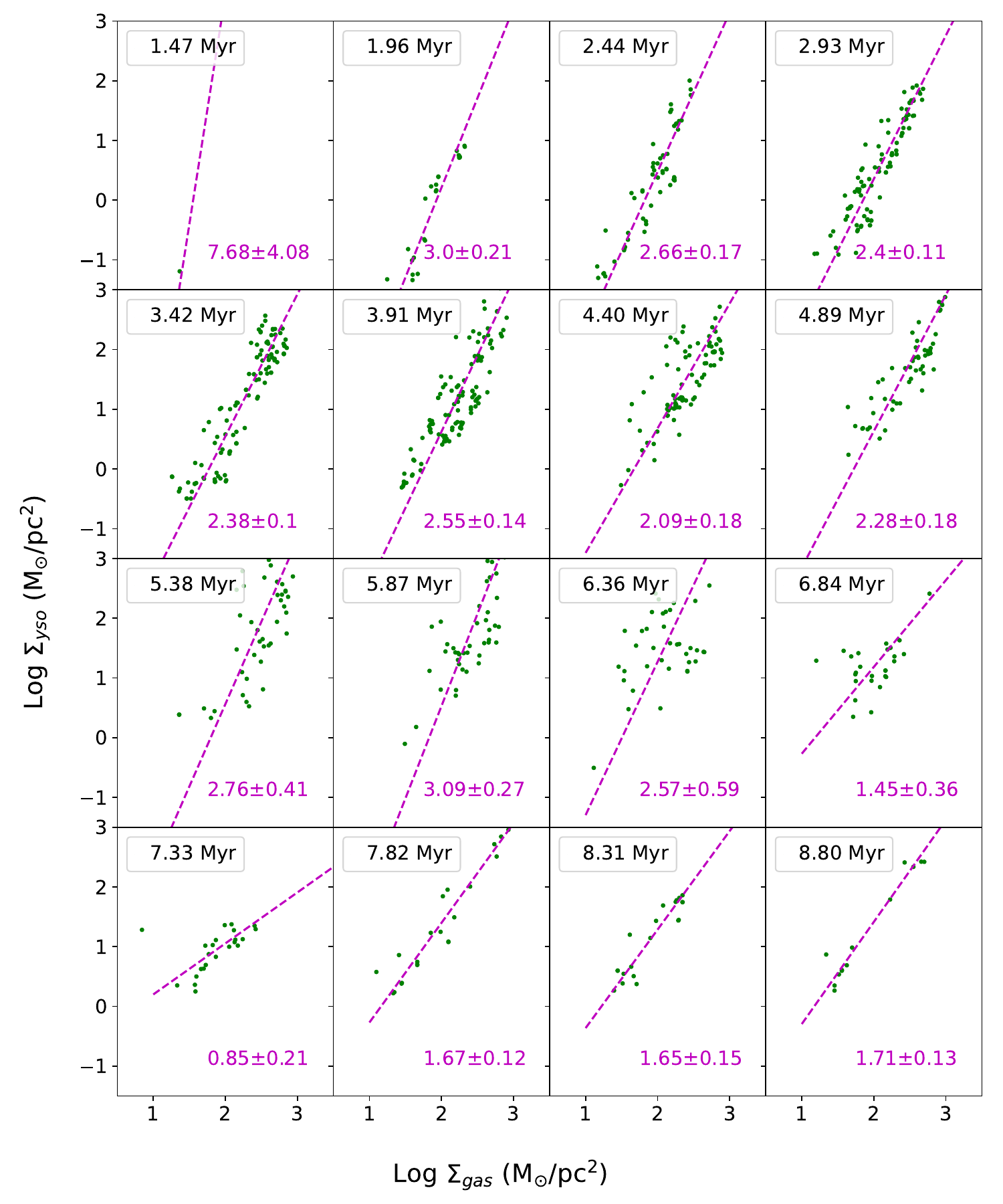}%
    \caption{S-G Correlation Plots for each snapshot in the \textbf{fiducial} analysis. Note that the first snapshot is extremely undersampled.}%
    \label{fig:S-G Default Overview}%
\end{figure*}

\begin{figure*}%
    \centering
    \includegraphics[width=.9\textwidth]{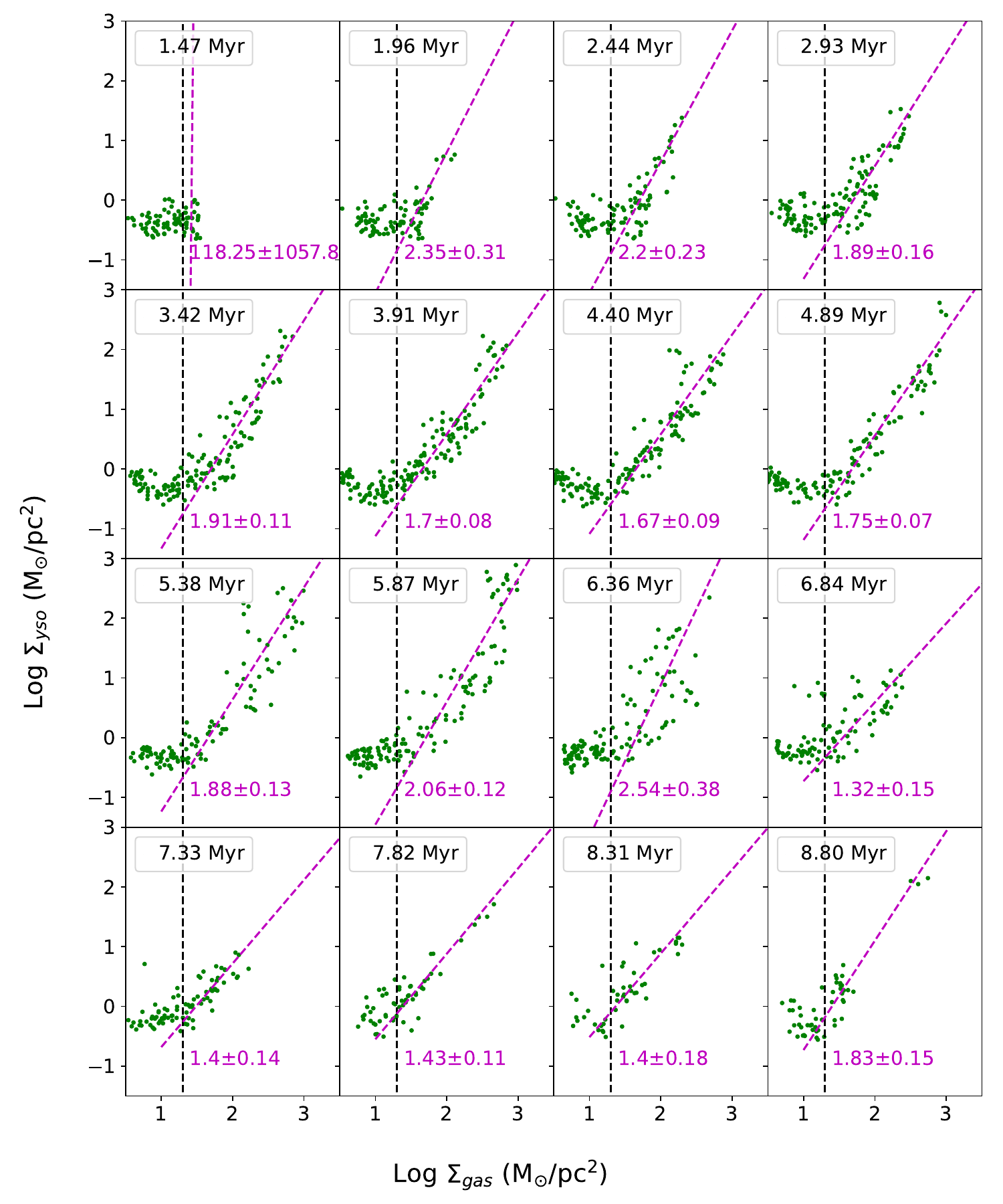}%
    \caption{S-G Correlation Plots for each snapshot in the \textbf{200 pc} synthetic observation run with all effects. This includes the removal of sources with $ {\rm log}(\Sigma_{\rm gas}) < 1.3 $ M$_{\odot}$/pc$^{2}$, represented by the black dotted line. Again, the first snapshot is undersampled. The clustered points in this snapshot are caused by AGN.}%
    \label{fig:S-G 200 Overview}%
\end{figure*}

\begin{figure*}%
    \centering
    \includegraphics[width=.9\textwidth]{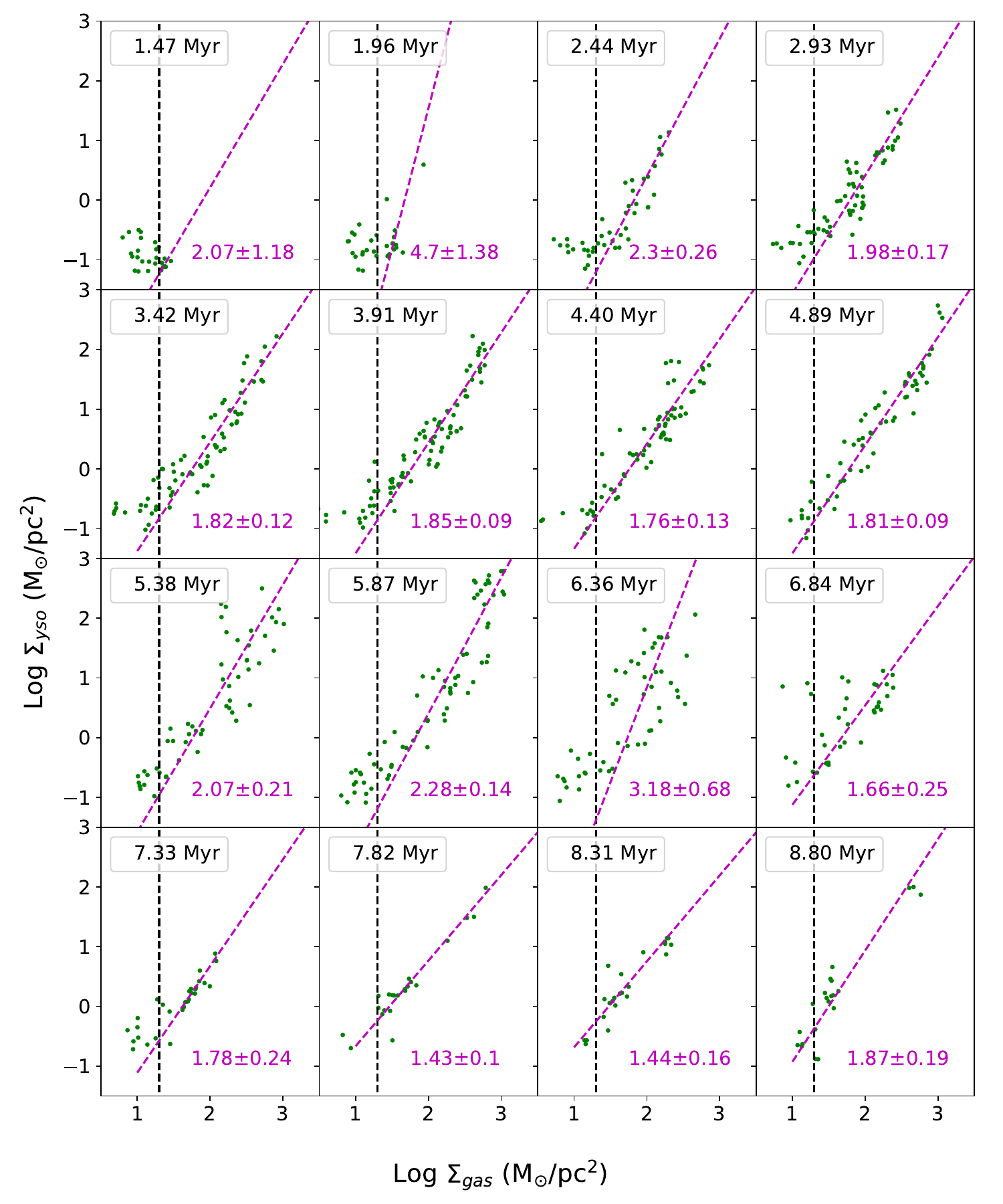}%
    \caption{S-G Correlation Plots for each snapshot in the \textbf{400 pc} synthetic observation run with all effects. This includes the removal of sources with $ {\rm log} (\Sigma_{\rm gas}) < 1.3 $ M$_{\odot}$/pc$^{2}$, represented by the black dotted line. As a result of observational effects, the early snapshots (especially those $<$ 2 Myr) are undersampled, consisting mostly of AGN that are at a greater gas density than the cutoff.}%
    \label{fig:S-G 400 Overview}%
\end{figure*}

\begin{figure*}%
    \centering
    \includegraphics[width=.9\textwidth]{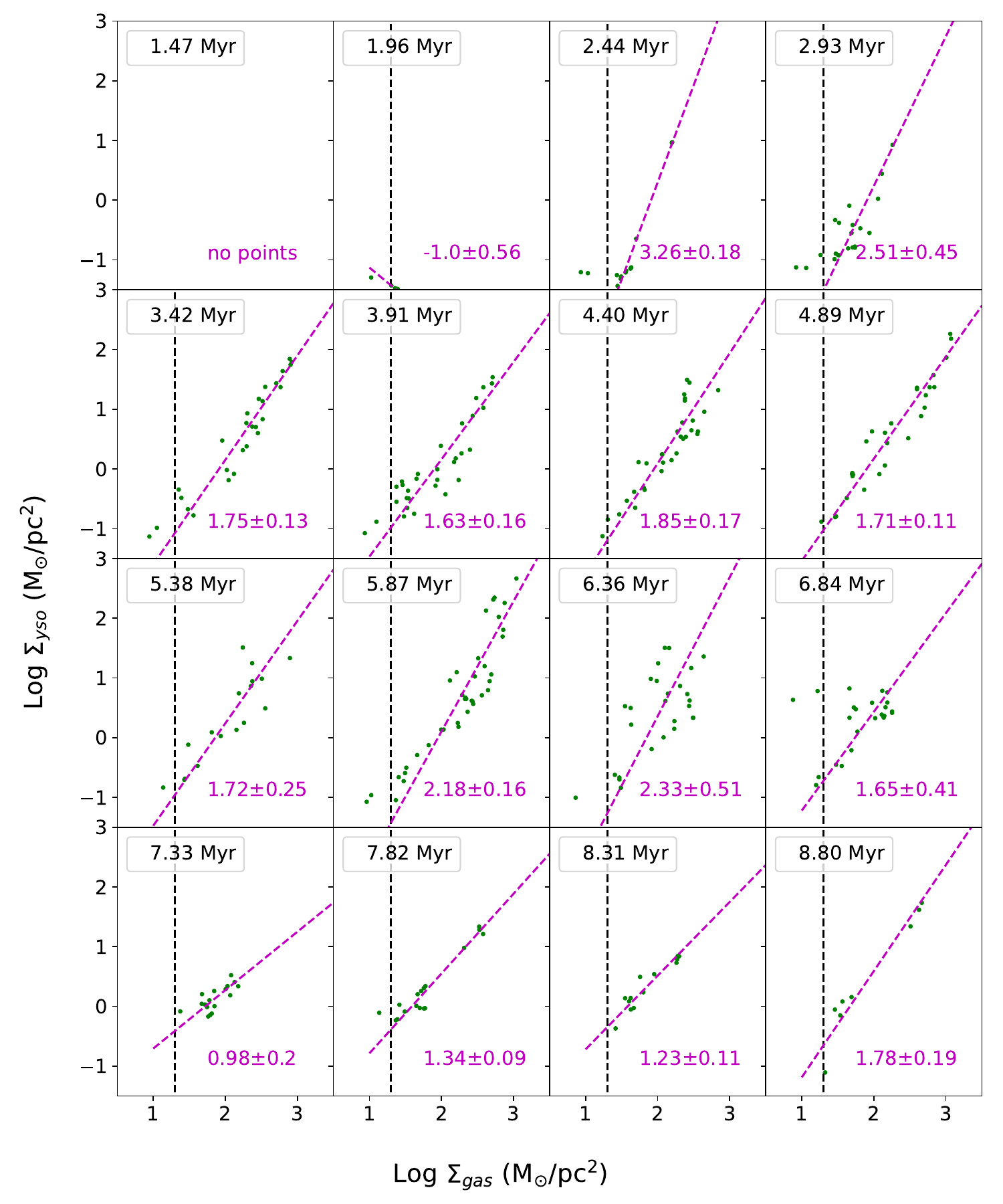}%
    \caption{S-G Correlation Plots for each snapshot in the \textbf{800 pc} synthetic observation run with all effects. This includes the removal of sources with log $\Sigma_{\rm gas} < 1.3 $ M$_{\odot}$/pc$^{2}$, represented by the black dotted line. Because of the relative lack of AGN and the increased mass and neighbor cutoffs, the first few snapshots are very undersampled. The 1.47 Myr snapshot does not even have 11 sources to create a single point on the S-G correlation plot.}%
    \label{fig:S-G 800 Overview}%
\end{figure*}

We present here in Figures \ref{fig:S-G Default Overview}, \ref{fig:S-G 200 Overview}, \ref{fig:S-G 400 Overview}, and \ref{fig:S-G 800 Overview} the full collection of S-G correlation plots for the fiducial and synthetic analyses. The fiducial case represents analysis done with minimal adjustments, while the others contain all synthetic effects described in Section \ref{section:Constructing Synthetic Observations}.

\label{lastpage}
\end{document}